\documentclass{aa}  

\usepackage{graphicx}
\usepackage{txfonts}


                              
\usepackage[colorlinks=true, linkcolor=blue, citecolor=blue, urlcolor=blue]{hyperref}
\usepackage{lipsum}

\newcommand{\msun}{M_{\odot}}

\newcommand{\be}{ \begin{equation} }
\newcommand{\ee}{\end{equation}}

\newcommand{\nb}{\texttt{NBODY6++GPU} }

\usepackage{xcolor}

\begin{document}

\title{Is the overconcentration of pristine populations in Galactic globular clusters real?}

\subtitle{An $N$-body approach to the problem}


   \author{P. Berczik\inst{2,3,1}
        \and O. Sobodar\inst{1,2}
        \and F. Flammini Dotti\inst{4,5,6}    
        \and M. Sobolenko\inst{1,2}
        \and M. Ishchenko\inst{1,2,3}
        \and R. Spurzem\inst{7,8,9}
        \and \goodbreak M. Giersz\inst{2}
        \and A. Askar\inst{2}
    }

   \institute{Main Astronomical Observatory, National Academy of Sciences of Ukraine, 27 Akademika Zabolotnoho St, 03143 Kyiv, Ukraine \\
   \email{\href{mailto:berczik@mao.kiev.ua}{berczik@mao.kiev.ua}}
   \and
   Nicolaus Copernicus Astronomical Centre, Polish Academy of Sciences, ul. Bartycka 18, 00-716 Warsaw, Poland
   \and 
   Fesenkov Astrophysical Institute, Observatory 23, 050020 Almaty, Kazakhstan
   \and
   Department of Physics, New York University Abu Dhabi, PO Box 129188 Abu Dhabi, UAE
   \and 
   Center for Astrophysics and Space Science (CASS), New York University Abu Dhabi, PO Box 129188, Abu Dhabi, UAE
   \and
   Dipartimento di Fisica, Sapienza, Universit\'a di Roma, P.le Aldo Moro, 5, 00185 - Rome, Italy
   \and 
   National Astronomical Observatories, Chinese Academy of Sciences, 20A Datun Rd., Chaoyang District, 100101, Beijing, China
   \and
   Kavli Institute for Astronomy and Astrophysics, Peking University, 5 Yi He Yuan Road, Beijing 100871, China
   \and
   Astronomisches Rechen-Institut, Zentrum f\"ur Astronomie der Universität Heidelberg, Mönchhofstraße 12-14, D-69120,  Germany
   }
   
\titlerunning{Pristine population overconcentration in globular clusters}
\authorrunning{Berczik et al.}

 
\abstract
{}
{Recent observations indicate that in some Milky Way globular clusters, pristine red giant branch (RGB) stars are more centrally concentrated than enriched ones. This contradicts most multiple stellar population (MSP) formation scenarios, which predict that the enriched (second) population (2P) should initially be more concentrated than the pristine (first) population (1P). Previous Monte Carlo Cluster Simulator (MOCCA) simulations suggested that this apparent overconcentration is a transient effect arising in clusters that have lost a large fraction of their initial mass and host an active black hole subsystem (BHS), and is visible only when RGB stars are used as tracers. We tested this interpretation using tailored \nb models evolved with direct $N$-body simulations and provide an independent validation that does not rely on a statistical treatment of relaxation.}
{We performed direct $N$-body simulations with the \nb code, adopting initial conditions designed to reproduce the dynamical regime relevant to the proposed mechanism. The simulations include updated stellar and binary evolution, dynamical interactions, and the Galactic tidal field, enabling a direct comparison with MOCCA results.}
{The simulations confirm that the spatial distributions and kinematics inferred from RGB stars can be strongly affected by stochastic fluctuations and interactions with the BHS. Preferential ejection of 2P RGB stars and their progenitors from the cluster centre leads to a transient apparent overconcentration of 1P RGB stars, in agreement with earlier MOCCA predictions. We show that this effect does not reflect the global MSP structure and that analyses based solely on RGB tracers may yield biased interpretations. These results support the view that dynamical evolution within the current MSP formation scenarios in our model can explain the apparent 1P overconcentration inferred in real clusters such as NGC~3201 and NGC~6101.}
{}

\keywords{stellar dynamics -- 
          methods: numerical -- 
          globular clusters: evolution -- 
          stars: multiple stellar populations}

\maketitle

\nolinenumbers

\section{Introduction}\label{sec:intro}

Globular clusters (GCs) are among the simplest stellar systems and have long been regarded as nearly spherical and chemically homogeneous. However, photometric and spectroscopic studies have revealed the presence of multiple stellar populations (MSPs), characterised by variations in light-element abundances (see \citealt{Bastian2018}, \citealt{Gratton2019}, and references therein). The pristine (first) population (1P) has a chemical composition similar to field stars of comparable metallicity, while the second population (2P) is enriched in light elements. The fraction of 2P stars in Milky Way GCs correlates strongly with GC mass, increasing from $\sim40\%$ in low-mass systems to $\sim90\%$ in the most massive clusters \citep{Milone2022}. 

Despite extensive efforts, the formation and evolution of MSPs remain uncertain. Most scenarios predict that the 2P forms more centrally concentrated than the 1P and gradually mixes with it through dynamical evolution \citep[e.g.][]{Bastian2018}. Observational studies largely support this idea and have characterised spatial and kinematic differences between sub-populations in increasing detail \citep[e.g.][]{Kamann2020,Kamann2020b,Libralato2023,Dalessandro2024,Leitingeretal2024,Cordoni2025}.  

Recently, however, \citet{Leitingeretal2023} reported that in NGC~3201 and NGC~6101 the 1P inferred from red giant branch (RGB) stars appears more centrally concentrated than 2P, challenging standard MSP scenarios. Subsequent work has shown that the interpretation of radial trends can depend sensitively on methodology and tracer selection. In particular, \citet{Cadelano2024} demonstrate that in NGC~3201 the enriched population remains more centrally concentrated within $\sim1.5\,r_{\rm h}$ and exhibits a bimodal radial structure not captured by cumulative diagnostics, while \citet{Mehta2025} independently confirmed the central concentration of enriched stars using \textit{Gaia} XP spectrophotometry. These results highlight the complexity of MSP radial structure and the importance of complementary diagnostics.

Motivated by these developments, \citet{Gierszetal2025} used Monte Carlo Cluster Simulator (MOCCA) simulations within the asymptotic giant branch (AGB) ejecta enrichment framework in which 2P stars form centrally from gas polluted by the processed ejecta of AGB stars. In this context, the `AGB ejecta enrichment framework' refers to models in which chemically enriched material from intermediate-mass AGB stars is retained within the cluster potential. Such scenarios were developed by \citet{Ventura2001, Ventura2013}, \citet{DErcole2008}, \citet{Renzini2015}, and \citet{Bastian2018}. A centrally concentrated 2P population is also a generic prediction of other enrichment models, including those involving fast-rotating massive stars \citep{Decressin2007} and supermassive stars \citep{Denissenkov2014}.

We tested this scenario using direct $N$-body simulations with the \nb code. As MOCCA relies on a statistical solution of the Fokker–Planck equations, independent verification with direct \textit{N}-body methods provides a complementary assessment of the proposed mechanism. By modelling clusters in the relevant dynamical regime, we examined whether the transient RGB-based inversion arises without Monte Carlo approximations. The paper is organised as follows. In Sect.~\ref{sec:method} we describe the \nb and MOCCA codes and outline the adopted initial conditions. In Sect.~\ref{sec:results} we present the simulation results. In Sect.~\ref{sec:concl} we summarise and discuss our findings.

\section{Method and initial conditions}\label{sec:method}
\subsection{The MOCCA and \nb code}\label{sec:mocca}

In this work we performed numerical simulations using the MOCCA code \citep{Giersz1998,Hypki2013,Gierszetal2013,Hypki2022,Hypki2025,Gierszetal2025b} and the direct $N$-body code \nb (\citealt{Spurzem2023} and references therein). MOCCA is based on Hénon’s Monte Carlo method \citep{Henon1971,Stodolkiewicz1982} and relies on a statistical solution of the Fokker–Planck equations. To model dynamical interactions of small-$N$ subsystems, it employs the \texttt{FEWBODY} package \citep{Fregeauetal2004,Fregeau2007}. MOCCA simulates the full stellar and dynamical evolution of realistic-size GCs up to a Hubble time.

\nb is a direct $N$-body code optimised for studying the dynamical evolution of star clusters and galaxies. It integrates large-$N$ systems using GPU acceleration and parallelisation across multiple GPU cores \citep{nitadori2012,Wangetal2015,HSB2016}. Recent developments of direct $N$-body frameworks for modelling composite stellar populations include NbodyCP \citep{LiSpurzem2026}, which highlights the growing capability of direct $N$-body codes to treat MSPs self-consistently. Hard binaries are handled using Kustaanheimo–Stiefel regularisation \citep{kustaanheimo1965}.

Both MOCCA and \nb use consistent prescriptions for single and binary stellar evolution, stellar winds, supernova kicks, and gravitational recoil from black hole mergers, with the most recent updates described in \citet{Kamlahetal2022} and \citet{Gierszetal2025b}. The MOCCA code has been extensively tested against direct $N$-body simulations, including \nb, and shows very good agreement across a wide range of cluster environments \citep[e.g.][]{Giersz2008,Wang2016,Madrid2017,Geller2019,Rizzuto2021,Kamlahetal2022,Vergara2025}.

\vspace*{-0.2cm}
\subsection{Initial conditions} \label{sec:init}

The initial conditions of the $N$-body model were chosen carefully, as large-scale direct $N$-body simulations are computationally expensive and time-consuming. Reproducing the cluster model presented in \citet{Gierszetal2025} would require months of computation, mainly because of its very high primordial binary fraction ($f_{\rm b} = 0.95$) and the non-parallelised treatment of internal binary evolution \citep{HSB2016}.  We therefore focused on reproducing the findings of \citet{Gierszetal2025}, namely the apparent 1P overconcentration in clusters close to dissolution, characterised by a small number of RGB stars and the presence of an active black hole subsystem (BHS). The selected $N$-body model dissolves on a much shorter timescale than a typical Milky Way GC, but it captures the essential physical processes relevant to the proposed mechanism.

The adopted initial conditions\footnote{Initial model and snapshot data are available at \url{https://zenodo.org/records/20070757}.} are as follows: initial number of particles $N_{\rm 1P} = 61\,000$ and $N_{\rm 2P} = 30\,500$, with a \citet{Kroupa2001} initial mass function. The 1P stars were sampled in the mass range $0.08$–$\rm 150\,M_{\odot}$, while the 2P stars were sampled between $0.08$ and $\rm 20\,M_{\odot}$. Both populations have metallicity $Z_{\rm 1P} = 0.001$ and $Z_{\rm 2P} = 0.001$, with binary fractions $f_{\rm b,1P} = 0.1$ and $f_{\rm b,2P} = 0.1$. These parameters correspond to an initial cluster mass of $M(0)=56\,953.8~\rm\msun$. The cluster was placed at a Galactocentric distance of $R_{\rm G} = 2$~kpc in a point-mass Galactic potential with $M_{\rm G} = 2.25 \times 10^{10}\rm \,M_{\odot}$. The ratio of half-mass radii is $R_{\rm h,2P}/R_{\rm h,1P} = 0.1$. The two populations are initially in virial equilibrium, with the 1P component slightly tidally under-filled ($R_{\rm tid} = 18.9$~pc). The initial King model \citep{King1966} concentration parameters were $W_0 = 3.0$ for 1P and $W_0 = 7.0$ for 2P.  The initial conditions were generated using the publicly available {\tt McLuster}\footnote{\url{https://github.com/agostinolev/mcluster}} code \citep{Kuepper2011,mcluster2022}.

\begin{figure}
\includegraphics[width=0.99\linewidth]{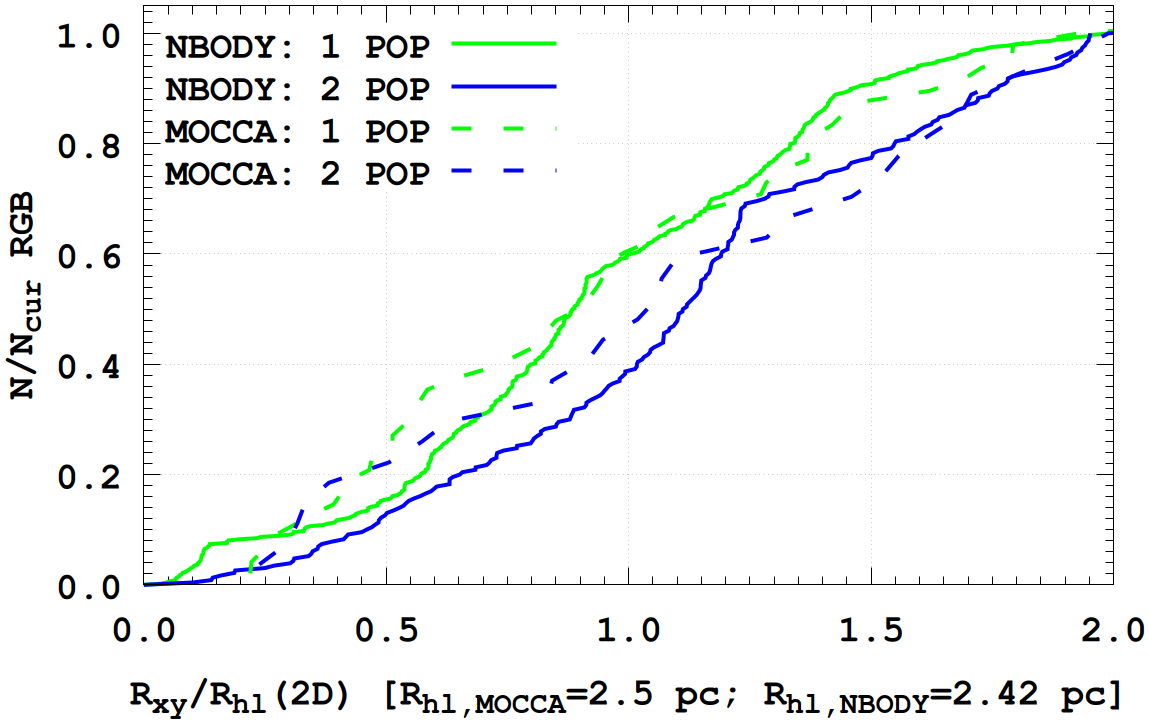}
\caption{Cumulative number distributions of RGB stars for the 1P (green) and 2P (blue) around 1.2~Gyr as a function of projected distance, normalised by the 2D half-light radius ($R_{\rm hl}$). The solid line represents the $N$-body model (average of 33 snapshots) and the dashed line the MOCCA model (average of 5 snapshots). Snapshots span 1.18--1.22~Gyr. More details are provided in Fig.~\ref{fig:N_vs_Rxy}.} 
\label{fig:def-Aplus}
\vspace*{-15pt}
\end{figure}

\section{Results}
\label{sec:results}

\begin{figure*}
\centering
\includegraphics[width=0.32\linewidth]{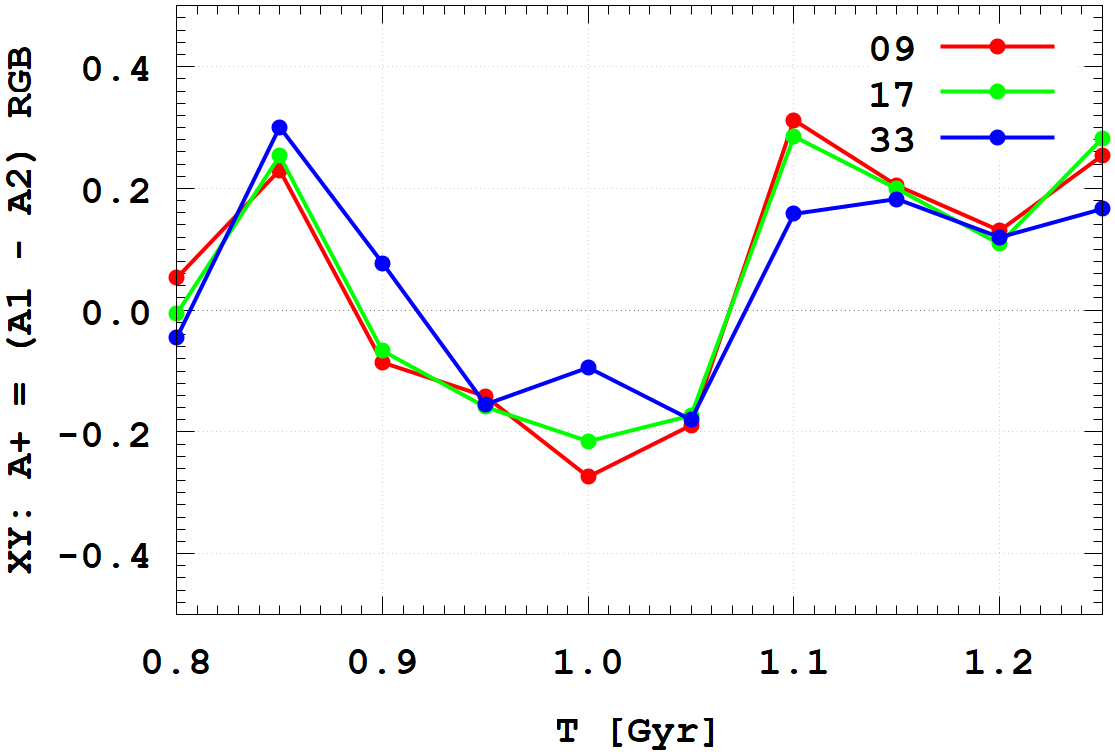}
\includegraphics[width=0.32\linewidth]{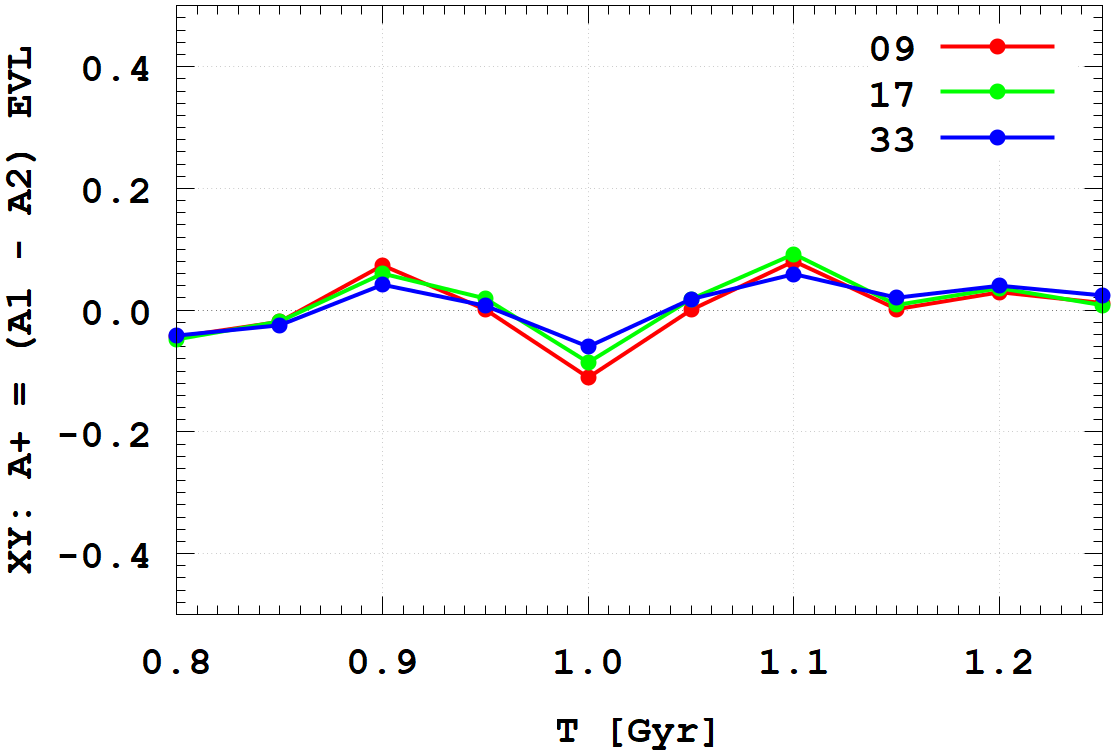}
\includegraphics[width=0.32\linewidth]{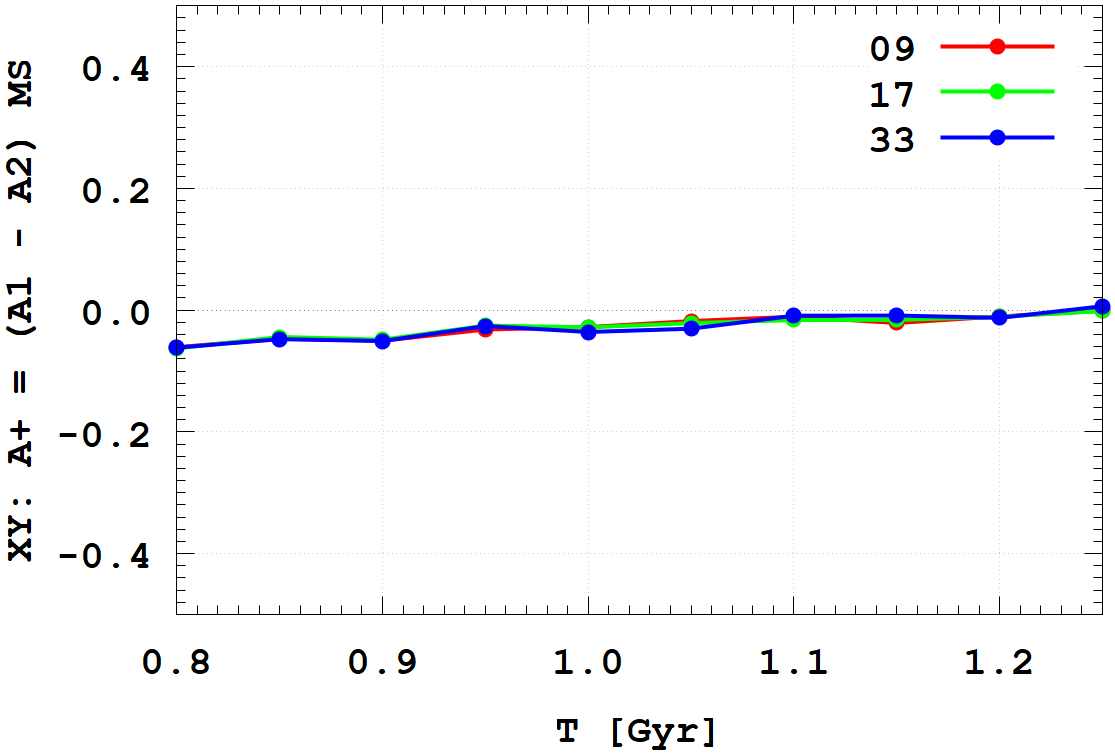}
\caption{Time evolution of the A$^+$ parameter for different types of stars and for the $X-Y$ projection of 3D snapshots. \textit{Left}: RGB stars. \textit{Middle}: Evolved luminous stars. \textit{Right}: MS stars. The red, green, and blue lines are averaged over 9, 17, and 33 snapshots, respectively.}
\label{fig:def-Aplus-01}
\end{figure*}

\begin{figure*}[hpbt!]
\centering
 \includegraphics[width=0.3\linewidth]{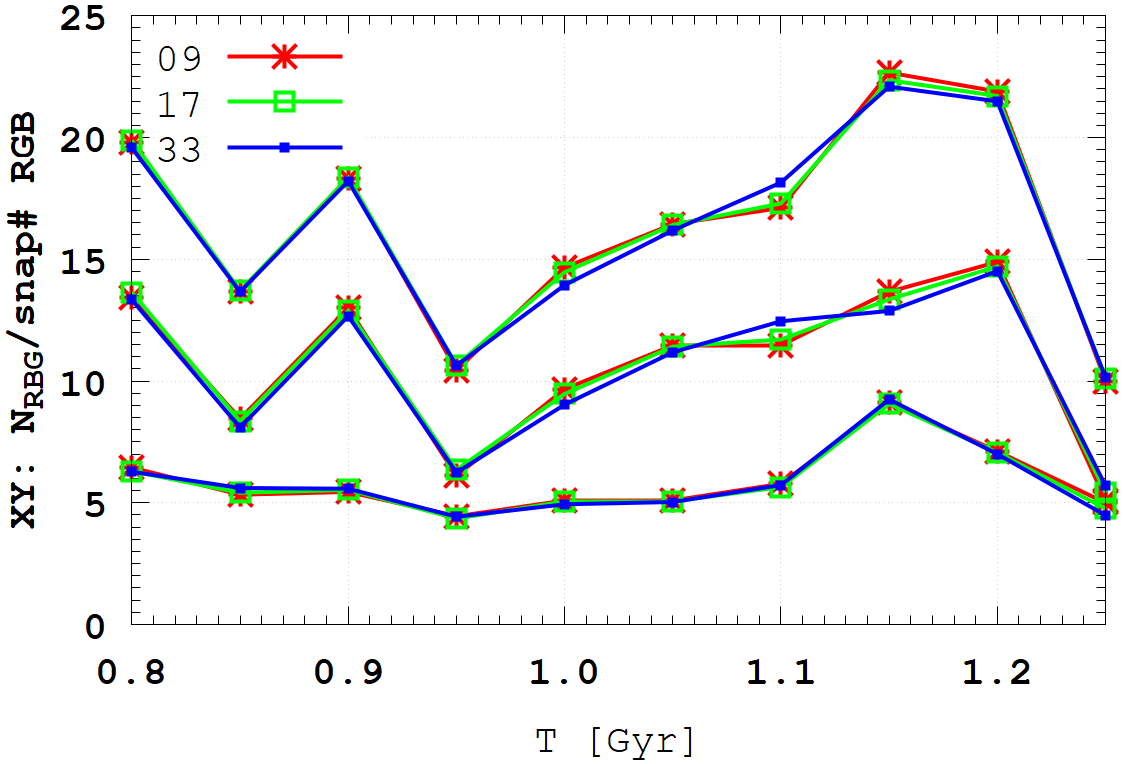}
\includegraphics[width=0.3\linewidth]{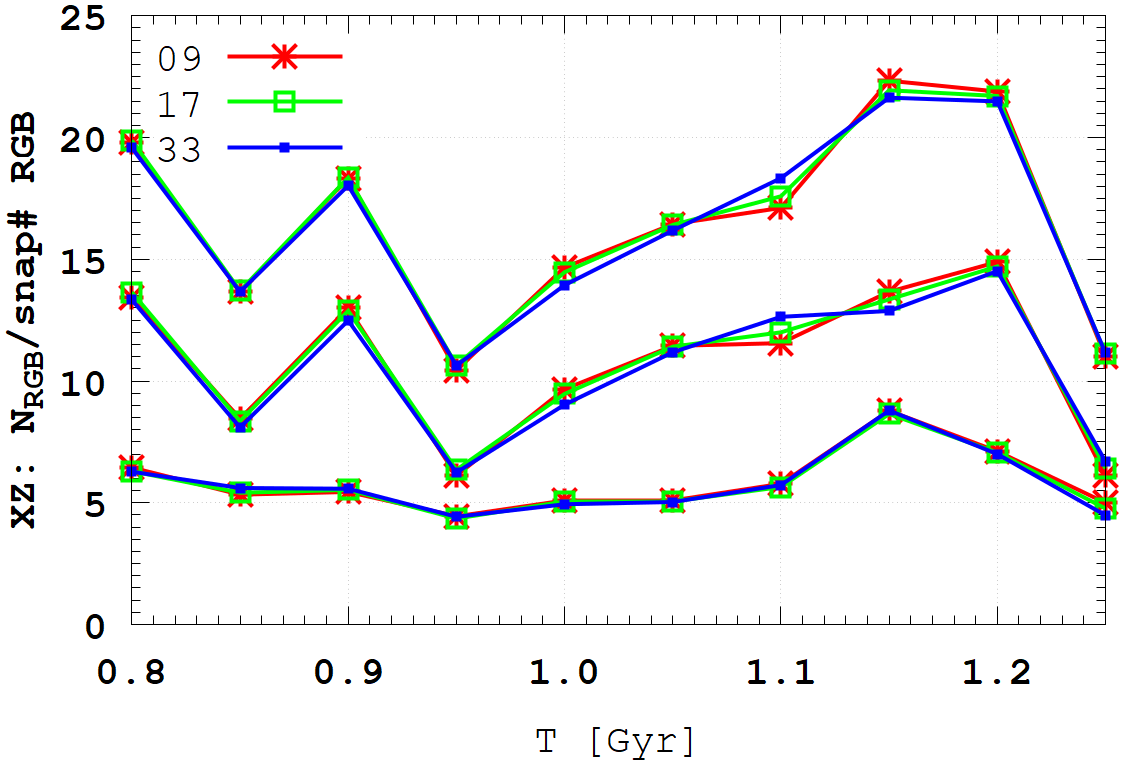}
\includegraphics[width=0.3\linewidth]{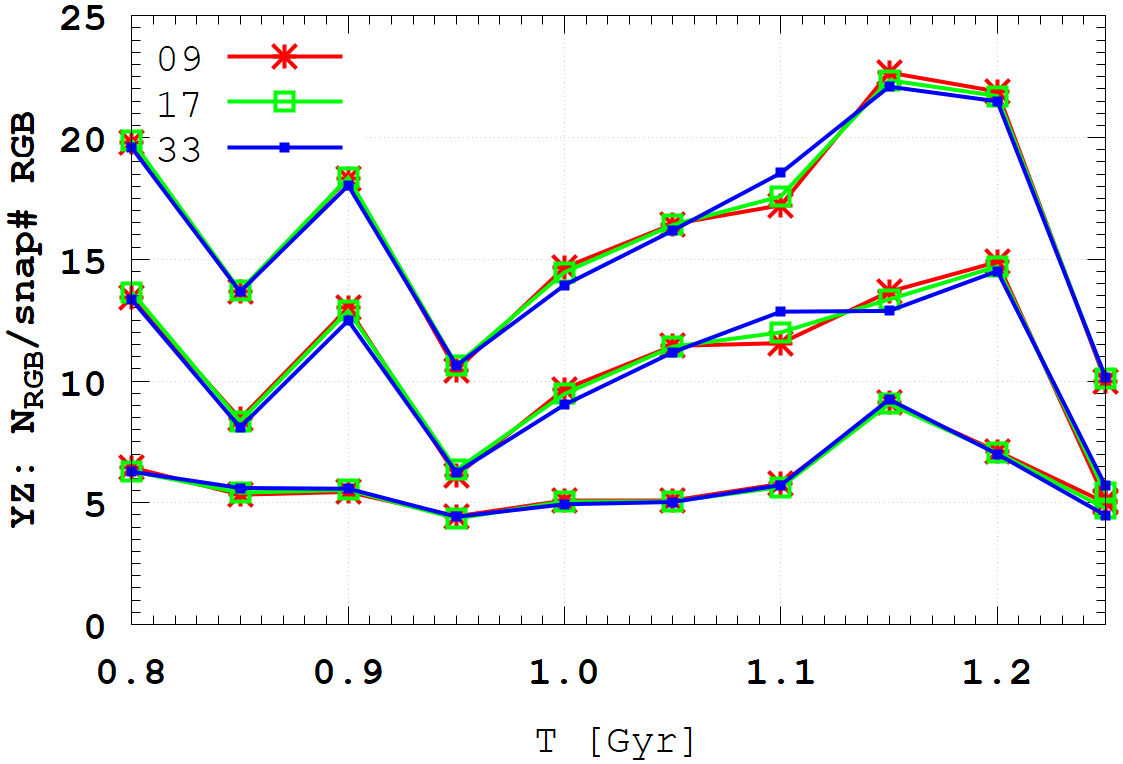}
\caption{Time evolution of the number of RGB stars in the 1P, 2P, and total population for different projections of the 3D snapshots. \textit{Left}: $X-Y$ projection. \textit{Middle}: $X-Z$ projection. \textit{Right}: $Y-Z$ projection. The red, green, and blue lines correspond to averages over 9, 17, and 33 snapshots centred on the selected time, respectively.}
\label{fig:def-Aplus-num}
\end{figure*}

Following the approach presented in \cite{Leitingeretal2023}, we used the A$^+$ parameter \citep{Alessandrini2016, Dalessandro2018, Dalessandro2019} to assess the degree of spatial mixing between the 1P and the 2P. At 1.2~Gyr we have $N_{\rm 1P} = 9\,021$ and $N_{\rm 2P} = 7\,206$. As an illustration of the A$^+$ parameter calculation, we show in Fig.~\ref{fig:def-Aplus} the cumulative projected distributions of RGB stars for both stellar populations of the $N$-body and MOCCA simulations. Both distributions are normalised by their numbers and their respective half-light radii. The total number of RGB stars at the reference time 1.2~Gyr is small, $N_{\rm RGB}$ = 22 (see Fig.~\ref{fig:def-Aplus-num}), so we collected data from snapshots between 1.18 and 1.22~Gyr and averaged the number distribution at 1.2~Gyr over a short time window. The MOCCA model has 5~snapshots, while the $N$-body model has 33~snapshots. The purpose of combining snapshots is not to increase the number of independent RGB stars, but to smooth the A+ measurement using a short time-smoothing window. 

Figure~\ref{fig:def-Aplus-num} shows the time evolution of the RGB-star counts in the 1P and 2P for projections of the 3D data onto the Galactic coordinate planes ($X-Y$, $X-Z$, and $Y-Z$). As expected, the RGB-star counts in each population are independent of projection direction and therefore show no systematic variation with projection. The number of RGB stars per snapshot is consistent and does not depend on the number of snapshots included in the averaging. The time steps are 0.116 Myr and 10 Myr for the \nb and MOCCA simulations, respectively.


\begin{figure*}[hpbt!]
\centering
\includegraphics[width=0.42\linewidth]{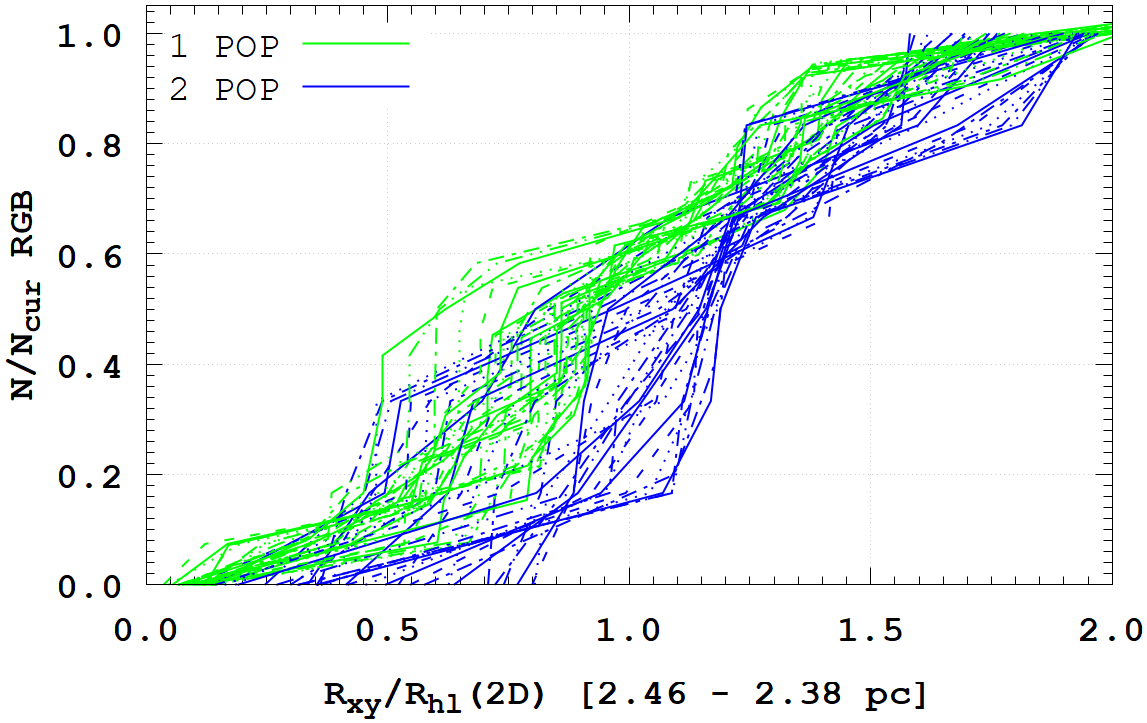}
\includegraphics[width=0.4\linewidth]{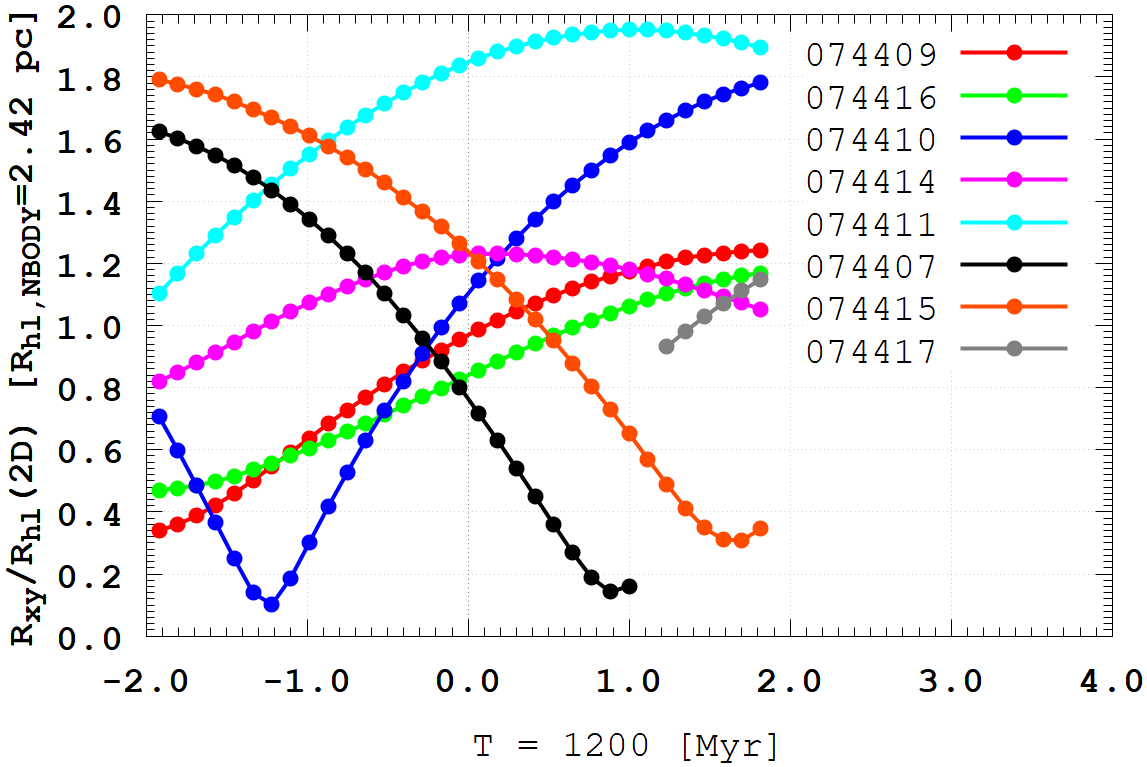}
\caption{\textit{Left}: Collection of the relative cumulative RGB star distribution for 1P and 2P stars for all 33 snapshots around 1.2 Gyr. Here we use for each snapshot its own half-light radius. These radii vary between 2.46 and 2.38 pc. \textit{Right}: Individual RGB stars' projected coordinates inside the cluster.}
\label{fig:N_vs_Rxy}
\end{figure*}

In Fig.~\ref{fig:N_vs_Rxy}, in the left panel, the data show the cumulative relative number distribution of RGB stars in both populations. Each line in the plot shows the individual cumulative distribution of RGB stars for the 33 separate snapshots inside a time interval of $\pm$2 Myr. The distribution fluctuates significantly between individual snapshots, making it extremely hard to determine the real A$^{+}$. The relative cumulative distribution changes from snapshot to snapshot reflect the RGB stars' physical movement inside the cluster, as shown in the right panel.  


In Fig.~\ref{fig:N_vs_Rxy}, in the right panel, we present the particle dynamical orbits (for the eight RGB stars of the 2P) around 1.2 Gyr. We see the projected orbits during the dynamical orbital evolution of these RGB stars inside the cluster. The coordinates of the RGB stars in these 33 snapshots change significantly, so the combination of these 33 snapshots into one `merged' system better represents the dynamical stage of the whole N-body system and the particle distribution. 

Figure~\ref{fig:def-Aplus} shows that both models predict the 1P RGB population to be more centrally concentrated than the 2P RGB population. Taking into account the fact that we are comparing two simulations done with two different codes, the agreement between the 2D projected mass distributions of RGB stars for the two models is very good. Small differences in the curves reflect the stochastic spatial distribution of the small RGB sample and the different time averaging of snapshots, since MOCCA averages over a longer interval than the $N$-body results; therefore, identical shapes are not expected. The A$^+$ parameters are 0.15 and 0.12 for the \nb and MOCCA models, respectively, at 1.2~Gyr. 

The time evolution of the A$^+$ parameter for different types of stars and for the $X-Y$ projection (in the Galactic coordinate system) of 3D snapshots is presented in Fig.~\ref{fig:def-Aplus-01}. The time range shown in the figure is from 0.8 to 1.25~Gyr. In the three panels, we show the time evolution for the different types of stars: RGB stars, evolved luminous stars (from the Hertzsprung gap to the helium giant branch), and main sequence (MS) stars. Each line presented in the figures was calculated by averaging a different number of snapshots. Each curve was computed using a sliding average over a window of snapshots centred on the reference time: 9 ($\pm$4), 17 ($\pm$8), or 33 ($\pm$16) snapshots. 

Generally, each of the three averages gives a more or less consistent result for MS stars (right panel) and other types of luminous stars (middle panel). For the most interesting case of RGB stars (left panel), the snapshot averaging gives slightly different results depending on the time, with a spread of at most about 0.1. The spread is much smaller than the variation in the A$^+$ parameter itself. Changes in the sign of the A$^+$ parameter occur on a timescale of about 0.15~Gyr (much longer than the averaging time), indicating a strong time variability. On short timescales, the 1P becomes overconcentrated relative to the 2P and then quickly returns to a less concentrated state. The 1P overconcentration is a transient feature and confirms the findings of \citet{Gierszetal2025}. 

The \citet{Gierszetal2025} conclusions are further confirmed by the time evolution of the A$^+$ parameter for MS and luminous evolved stars. No significant variation is seen in the A$^+$ parameter, which remains close to 0. Both populations are fully mixed for MS and evolved luminous stars. 

\begin{figure*}[hpbt!]
\centering
 \includegraphics[width=0.3\linewidth]{pic/def-Aplus-new-03.XY.png}
\includegraphics[width=0.3\linewidth]{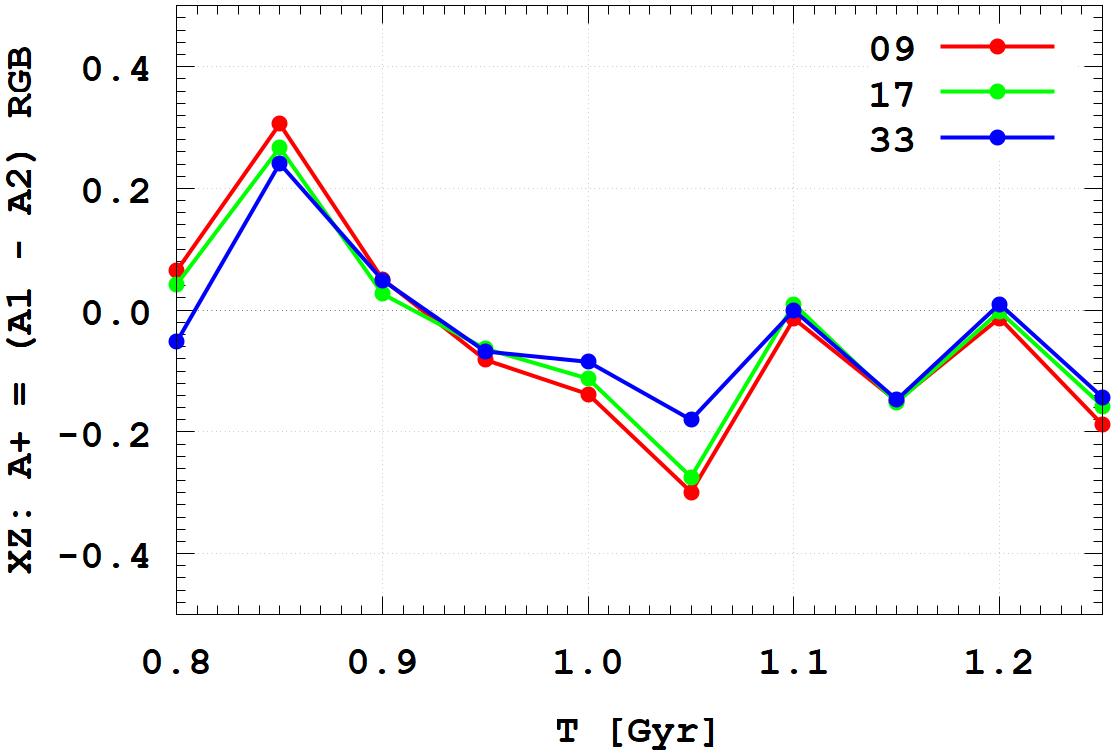}
\includegraphics[width=0.3\linewidth]{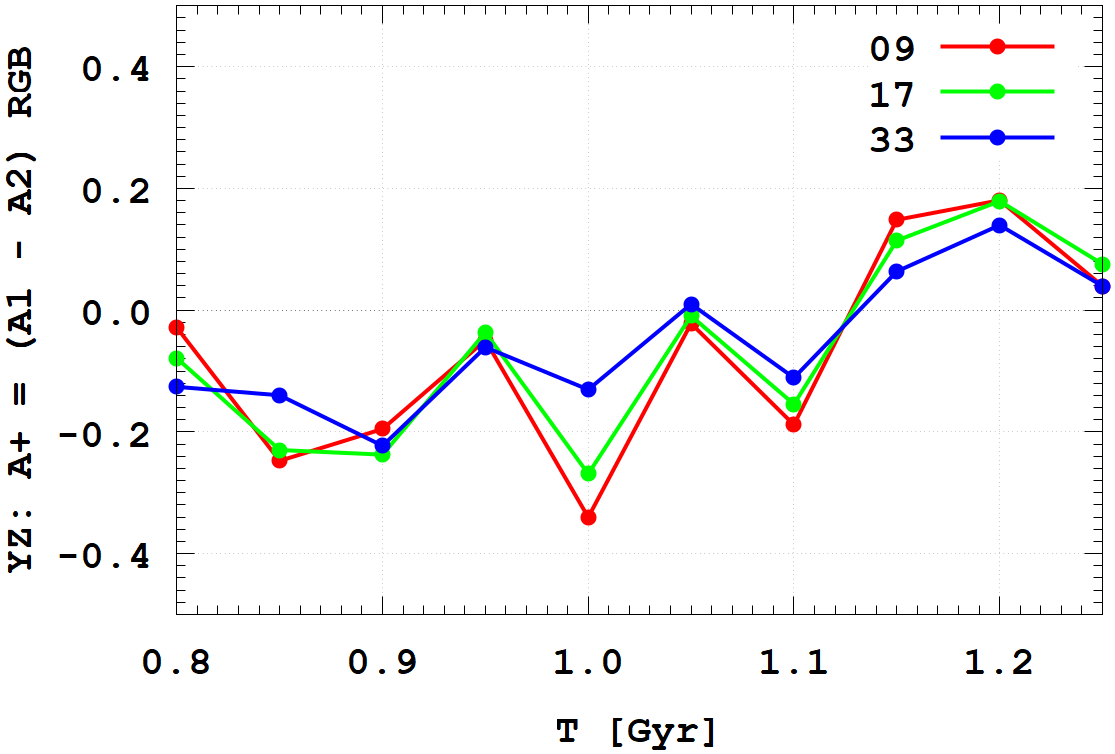}
\caption{Time evolution of the A$^+$ parameter for RGB stars for projections of the 3D snapshots onto the $X-Y$, $X-Z$, and $Y-Z$ planes (\textit{left to right}). The red, green, and blue lines show averages over 9, 17, and 33 snapshots, respectively.}
\label{fig:def-Aplus-yz}
\end{figure*}

In Fig.~\ref{fig:def-Aplus-yz} we show the evolution of the A$^+$ parameter for different projections of the 3D snapshots onto the $X- Y$, $X-Z$, and $Y-Z$ planes. The behaviour of A$^+$ differs between projections, although the amplitude of the fluctuations remains comparable. At 1.2~Gyr, A$^+$ is positive for the $X- Y$ and $Y-Z$ projections, while it is close to zero for the $X-Z$ projection. This indicates that the inferred spatial distribution of RGB stars depends on the viewing direction, suggesting a non-spherical distribution of RGB stars. A weak dependence of A$^+$ on the number of averaged snapshots is also present. The statistical fluctuations associated with different averaging windows are at the level of $\pm 0.05-0.1$, which is significantly lower than the intrinsic variations in A$^+$ occurring over time intervals of 0.1--0.15~Gyr. This confirms that the transient behaviour of A$^+$ reflects genuine changes in the global spatial distribution of RGB stars rather than artefacts of snapshot averaging.

Further confirmation of the transient feature of the overconcentration of the 1P RGB stars compared to the 2P RGB stars is provided by Fig.~\ref{fig:def-Aplus-yz}, in which the evolution of the A$^+$ parameter is presented for different projections of the 3D snapshots: $X-Y$, $X-Z$, and $Y-Z$ projections. We see that the A$^+$ parameter shows similar strong variations and a transient character for all projections. The transient overconcentration of 1P RGB stars is not driven by projection effects but instead reflects the time evolution of the spatial distributions of RGB stars in the 1P and 2P.

\begin{figure}[hpbt!]
\centering
\includegraphics[width=0.99\linewidth]{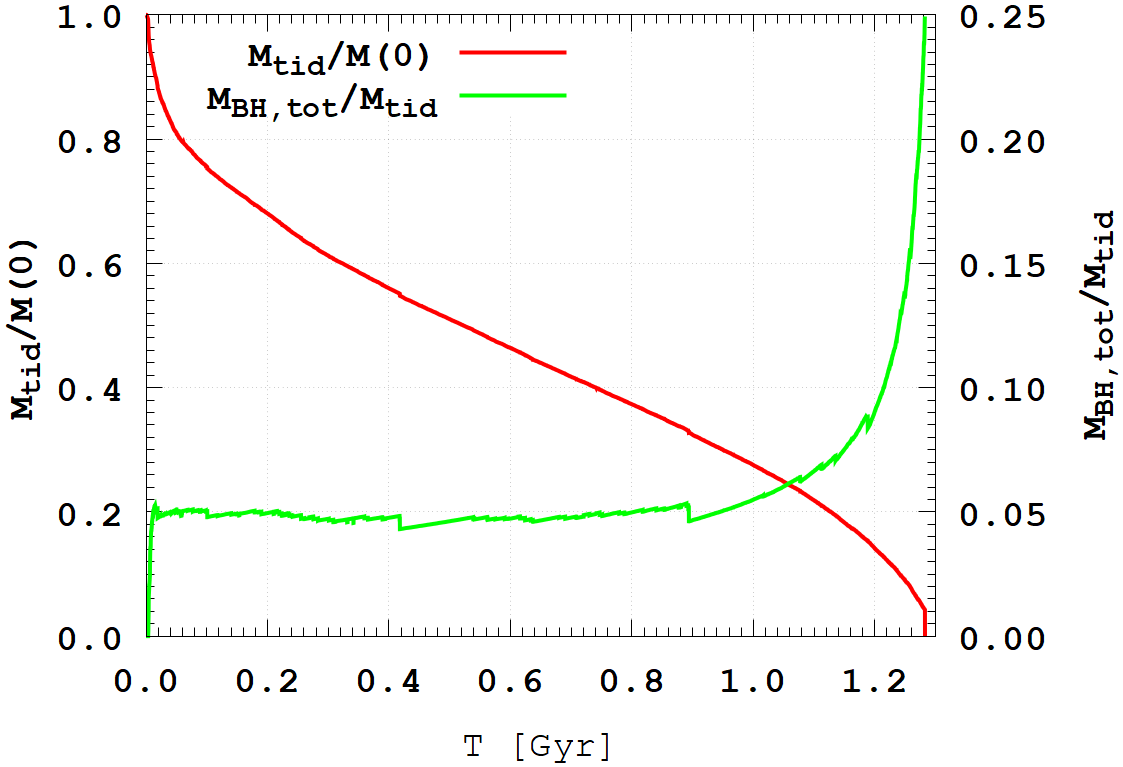}
\caption{\nb evolution of the cluster mass normalised to its initial value, $M(0) = 56953.8\rm\;M_{\odot}$ (red line), and of the BHS mass normalised to the current cluster mass (green line).}
\label{fig:bh-mass3}
\end{figure}

In Fig.~\ref{fig:bh-mass3} we show the evolution of the cluster mass normalised to its initial value, together with the ratio of the BHS mass to the current cluster mass. This figure illustrates the key cluster state associated with the transient behaviour of the A$^+$ parameter. When the cluster mass decreases to $\sim$ 0.1 of its initial value and the BHS mass reaches $\sim$ 0.05 of the current cluster mass, the number of RGB stars becomes very small. In this regime, stochastic fluctuations in their spatial distribution and strong dynamical interactions with the BHS drive the transient variations in A$^+$. This behaviour is fully consistent with the findings of \citet{Gierszetal2025}. 

\section{Conclusions}\label{sec:concl}

We confirm the results of \citet{Gierszetal2025}, showing that when the analysis is based on RGB stars, the 1P can appear more centrally concentrated than the 2P at radii of a few half-light radii. Using direct $N$-body simulations tailored to reproduce the dynamical regime relevant to the proposed mechanism, we demonstrate that this apparent inversion is a transient feature and depends strongly on the adopted stellar tracers.

This effect is particularly relevant for GCs with present-day masses of a few times $\rm 10^{5}\,\msun$, which have retained only $\sim$10\% of their initial mass and host an active BHS. In such systems, small-number statistics and strong dynamical interactions can significantly influence the spatial distribution of RGB stars, potentially leading to biased inferences about the global MSP structure when observations are restricted to this tracer population. To capture these processes while maintaining a feasible computational cost, we adopted a reduced-$N$ model with an accelerated evolutionary timescale. Despite these simplifications, the simulations reproduce the key evolutionary behaviour of the reference MOCCA model. In line with \citet{Gierszetal2025}, the simulated cluster hosts an active BHS and is close to dissolution, with black holes contributing $\sim$8\% of the cluster mass when only $\sim$10\% of the initial mass remains, as seen in Fig.~\ref{fig:bh-mass3}.

Figure\,~\ref{fig:def-Aplus-01} shows that the A$^+$ parameter derived from RGB stars varies strongly over time, ranging between $-0.3$ and $0.4$ over $\sim$0.45~Gyr. In contrast, for evolved luminous stars (excluding compact remnants), A$^+$ remains close to zero, indicating a near-complete mixing of the populations. When computed using only MS stars, the evolution follows the expected monotonic trend towards spatial mixing. The MOCCA simulation exhibits a comparable transient signal near 1.2~Gyr, with A$^+ \simeq$ 0.12, in close agreement with the \nb result (A$^+ \simeq$ 0.15), providing an independent validation of the Monte Carlo interpretation. 

\begin{figure}[hpbt!]
\centering
\includegraphics[width=0.99\linewidth]{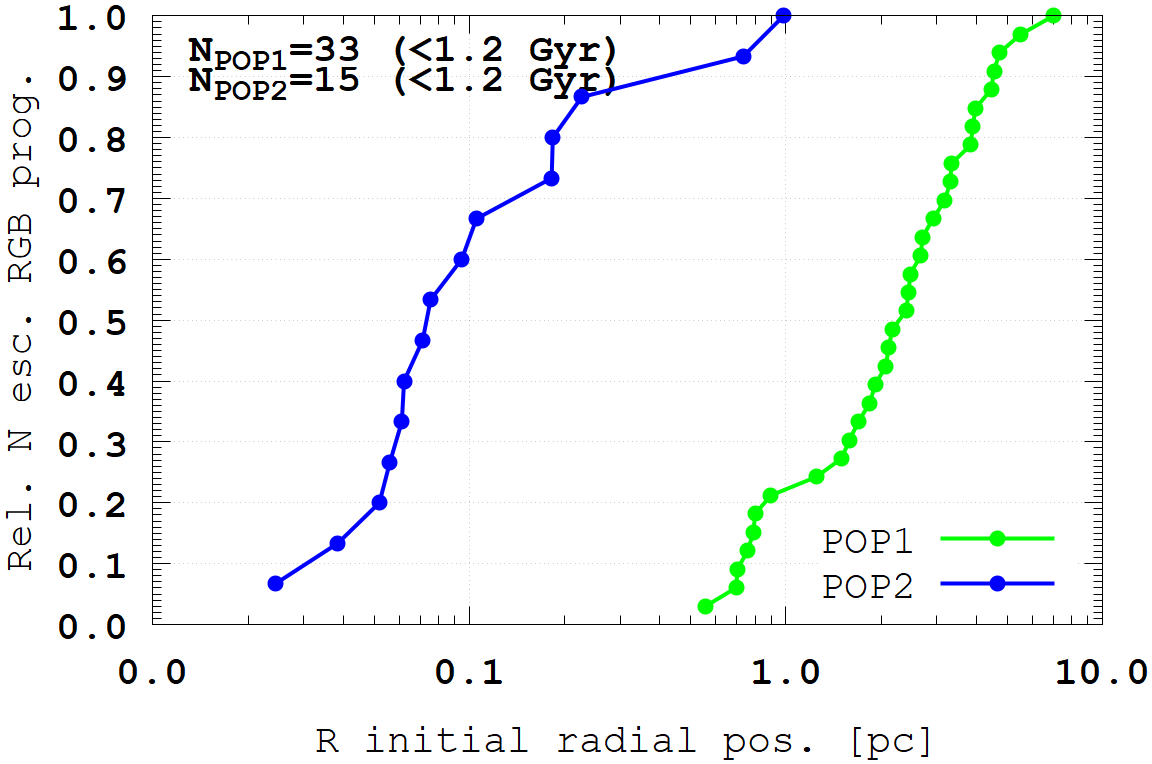}
\caption{Cumulative distributions of the initial radial positions of stars with initial masses in the range 1.71–1.74 $\rm M_{\odot}$ that would have evolved into RGB stars by 1.2 Gyr but escaped the cluster prior to this time. The green curve shows 1P escapers (N = 33 out of a total of 77 such stars in the initial model), and the blue curve shows 2P escapers (N = 15 out of a total of 40). The distributions demonstrate that 2P escapers originate predominantly from the inner regions of the cluster, whereas 1P escapers are drawn from a much broader radial range. This provides direct evidence that centrally concentrated 2P RGB progenitors are preferentially removed from the central parts of the cluster.}
\label{fig:def-esc-pop}
\end{figure}

To further investigate the origin of the RGB trends discussed above and the cumulative number distributions shown in Fig.~\ref{fig:def-Aplus}, we examined the escape properties of stars that would have evolved into RGB stars at 1.2 Gyr. In particular, we selected stars with initial masses in the range 1.71--1.74~$\rm M_{\odot}$ and identified those that escaped the cluster prior to this time. In Fig.~\ref{fig:def-esc-pop} we show the cumulative distributions of their initial radial positions for the two populations, considering only escapers with securely identified population tags. As seen, the majority of 2P escapers originate from the innermost regions of the cluster, whereas 1P escapers are drawn from a much broader radial range. This demonstrates that centrally concentrated 2P RGB progenitors are preferentially removed through dynamical interactions in the cluster core, consistent with the behaviour shown in Fig.~3 of \citet{Gierszetal2025}. We also note that the number of escaped RGB progenitors is comparable to the number of such stars remaining in the cluster at 1.2 Gyr, indicating that this depletion is dynamically significant. We note that the number of identified 2P escapers should be regarded as a lower limit since a fraction of escapers cannot be unambiguously assigned to a given population. Nevertheless, the trend remains clear when considering only securely identified objects.

A similar role of dynamical interactions in removing centrally concentrated 2P stars from the cluster centre has recently been identified by \citet{pavlik2025}, who investigated spatial mixing driven by binary–single scattering with direct $N$-body simulations. Although their models do not show an inversion of the cumulative profiles, enriched stars are preferentially scattered to larger radii through binary–single encounters. The absence of inversion likely reflects the lack of strong mass loss and the near-dissolution evolution in their clusters. Taken together, these results suggest that while dynamical scattering can significantly modify MSP radial structure, the emergence of an apparent inversion requires clusters to have lost a substantial fraction of their initial mass. In this context, GCs such as NGC~3201 and NGC~6101, where 1P RGB stars are observed to be spatially overconcentrated, provide an interesting comparison. While these systems are inferred to have relatively high present-day mass fractions (e.g. \citealt{Baumgardt2003,Leitingeretal2023}), such estimates rely on simplified assumptions about their orbital history and tidal environment (see the discussion in \citealt{Gierszetal2025}) and may therefore be uncertain.

Projection-dependent variations in the inferred A$^+$ evolution further demonstrate the sensitivity of cumulative diagnostics to sampling effects and viewing geometry. While these fluctuations do not alter the qualitative conclusion regarding the transient nature of RGB-based segregation, they highlight the importance of using multiple complementary diagnostics when interpreting spatial distributions. This is particularly relevant in light of recent observational work showing that MSPs have a complex and sometimes non-monotonic radial behaviour \cite[e.g.][]{Cadelano2024,Mehta2025}, and that cumulative metrics alone may not fully capture underlying structural features. Our results, therefore, do not imply that enriched populations are globally less concentrated than pristine ones, but rather that tracer selection and the dynamical state can produce apparent inversions in specific diagnostics. In this context, the transient behaviour identified here may represent a dynamical pathway contributing to the observational diversity reported in recent studies. 

We conclude that analyses based solely on RGB tracers should be interpreted with caution and that complementary approaches that use larger and dynamically less biased samples, such as MS stars, are essential for a robust inference of MSP structure. Further observational tests of tracer-dependent segregation and comparisons with diagnostics beyond cumulative distributions would provide valuable constraints on MSP formation scenarios, including the AGB framework and alternatives.

\begin{acknowledgements}

The authors thank Sebastian Kamann for a very helpful and constructive comments. 
PB, MI, MS and OS thank the support from the special program of the Polish Academy 
of Sciences and the U.S. National Academy of Sciences under the Long-term program 
to support Ukrainian research teams, grant No.~PAN.BFB.S.BWZ.329.022.2023. 
The authors appreciate the Polish high-performance computing infrastructure PLGrid 
(HPC Centre: ACK Cyfronet AGH -- docs.hpc.cyfronet.pl) for providing Helios computer 
facilities and support within computational grant No.~PLG/2026/019243. The authors 
also acknowledge the Gauss Centre for Supercomputing e.V. (www.gauss-centre.eu) 
for funding this project by providing computing time through the John von Neumann 
Institute for Computing (NIC) on the GCS Supercomputer JUWELS Booster and JUPITER 
Booster at Julich Supercomputing Centre (JSC), Germany. This material is based upon 
work supported by Tamkeen under the NYU Abu Dhabi Research Institute grant CASS. 
PB and MI thank Project No. BR24992759 ``Development of the concept for the first 
Kazakhstani orbital cislunar telescope - Phase I'', financed by the Ministry of 
Science and Higher Education of the Republic of Kazakhstan. RS acknowledges NAOC 
International Cooperation Office for its support in 2023, 2024, and 2025. 
RS acknowledges Chinese Academy of Sciences President's International Fellowship 
Initiative for Visiting Scientists (PIFI, grant No. 2026PVA0089), and the 
National Natural Science Foundation of China (NSFC) under grant No. 12473017. 
This research was supported in part by the grant NSF PHY-2309135 to the 
Kavli Institute for Theoretical Physics (KITP). 
RS and FFD acknowledge German Science Foundation (DFG) grant Sp 345/24-1. 
MG was supported by the Polish National Science Centre (NCN) through the grant 2021/41/B/ST9/01191. 
AA acknowledges that this research was funded in part by National 
Science Centre (NCN), Poland, grant No.~2024/55/D/ST9/02585. 
For the purpose of Open Access, the author has applied a
CC BY public copyright licence to any Author Accepted
Manuscript (AAM) version arising from this submission.

\end{acknowledgements}
\bibliographystyle{bibtex/aa}  
\bibliography{bibtex/sources.bib}   

@ARTICLE{Denissenkov2014,
       author = {{Denissenkov}, P.~A. and {Hartwick}, F.~D.~A.},
        title = "{Supermassive stars as a source of abundance anomalies of proton-capture elements in globular clusters}",
      journal = {\mnras},
         year = 2014,
        month = jan,
       volume = {437},
       number = {1},
        pages = {L21-L25},
          doi = {10.1093/mnrasl/slt133},
archivePrefix = {arXiv},
       eprint = {1305.5975},
 primaryClass = {astro-ph.SR},
       adsurl = {https://ui.adsabs.harvard.edu/abs/2014MNRAS.437L..21D}
}

@ARTICLE{Decressin2007,
       author = {{Decressin}, T. and {Meynet}, G. and {Charbonnel}, C. and {Prantzos}, N. and {Ekstr{\"o}m}, S.},
        title = "{Fast rotating massive stars and the origin of the abundance patterns in galactic globular clusters}",
      journal = {\aap},
         year = 2007,
        month = mar,
       volume = {464},
       number = {3},
        pages = {1029-1044},
          doi = {10.1051/0004-6361:20066013},
archivePrefix = {arXiv},
       eprint = {astro-ph/0611379},
 primaryClass = {astro-ph},
       adsurl = {https://ui.adsabs.harvard.edu/abs/2007A&A...464.1029D}
}

@ARTICLE{Renzini2015,
       author = {{Renzini}, A. and {D'Antona}, F. and {Cassisi}, S. and {King}, I.~R. and {Milone}, A.~P. and {Ventura}, P. and {Anderson}, J. and {Bedin}, L.~R. and {Bellini}, A. and {Brown}, T.~M. and {Piotto}, G. and {van der Marel}, R.~P. and {Barbuy}, B. and {Dalessandro}, E. and {Hidalgo}, S. and {Marino}, A.~F. and {Ortolani}, S. and {Salaris}, M. and {Sarajedini}, A.},
        title = "{The Hubble Space Telescope UV Legacy Survey of Galactic Globular Clusters - V. Constraints on formation scenarios}",
      journal = {\mnras},
         year = 2015,
        month = dec,
       volume = {454},
       number = {4},
        pages = {4197-4207},
          doi = {10.1093/mnras/stv2268},
archivePrefix = {arXiv},
       eprint = {1510.01468},
 primaryClass = {astro-ph.GA},
       adsurl = {https://ui.adsabs.harvard.edu/abs/2015MNRAS.454.4197R}
}

@ARTICLE{DErcole2008,
       author = {{D'Ercole}, Annibale and {Vesperini}, Enrico and {D'Antona}, Francesca and {McMillan}, Stephen L.~W. and {Recchi}, Simone},
        title = "{Formation and dynamical evolution of multiple stellar generations in globular clusters}",
      journal = {\mnras},
         year = 2008,
        month = dec,
       volume = {391},
       number = {2},
        pages = {825-843},
          doi = {10.1111/j.1365-2966.2008.13915.x},
archivePrefix = {arXiv},
       eprint = {0809.1438},
 primaryClass = {astro-ph},
       adsurl = {https://ui.adsabs.harvard.edu/abs/2008MNRAS.391..825D}
}

@ARTICLE{Ventura2013,
       author = {{Ventura}, P. and {Di Criscienzo}, M. and {Carini}, R. and {D'Antona}, F.},
        title = "{Yields of AGB and SAGB models with chemistry of low- and high-metallicity globular clusters}",
      journal = {\mnras},
         year = 2013,
        month = jun,
       volume = {431},
       number = {4},
        pages = {3642-3653},
          doi = {10.1093/mnras/stt444},
archivePrefix = {arXiv},
       eprint = {1303.3912},
 primaryClass = {astro-ph.SR},
       adsurl = {https://ui.adsabs.harvard.edu/abs/2013MNRAS.431.3642V}
}

@ARTICLE{Ventura2001,
       author = {{Ventura}, Paolo and {D'Antona}, Francesca and {Mazzitelli}, Italo and {Gratton}, Raffaele},
        title = "{Predictions for Self-Pollution in Globular Cluster Stars}",
      journal = {\apjl},
         year = 2001,
        month = mar,
       volume = {550},
       number = {1},
        pages = {L65-L69},
          doi = {10.1086/319496},
archivePrefix = {arXiv},
       eprint = {astro-ph/0103337},
 primaryClass = {astro-ph},
       adsurl = {https://ui.adsabs.harvard.edu/abs/2001ApJ...550L..65V}
}

@ARTICLE{Dalessandro2018,
       author = {{Dalessandro}, E. and {Cadelano}, M. and {Vesperini}, E. and {Salaris}, M. and {Ferraro}, F.~R. and {Lanzoni}, B. and {Raso}, S. and {Hong}, J. and {Webb}, J.~J. and {Zocchi}, A.},
        title = "{The Peculiar Radial Distribution of Multiple Populations in the Massive Globular Cluster M80}",
      journal = {\apj},
         year = 2018,
        month = may,
       volume = {859},
       number = {1},
          eid = {15},
        pages = {15},
          doi = {10.3847/1538-4357/aabb56},
archivePrefix = {arXiv},
       eprint = {1804.03222},
 primaryClass = {astro-ph.SR},
       adsurl = {https://ui.adsabs.harvard.edu/abs/2018ApJ...859...15D}
}

@ARTICLE{Dalessandro2019,
       author = {{Dalessandro}, Emanuele and {Cadelano}, M. and {Vesperini}, E. and {Martocchia}, S. and {Ferraro}, F.~R. and {Lanzoni}, B. and {Bastian}, N. and {Hong}, J. and {Sanna}, N.},
        title = "{A Family Picture: Tracing the Dynamical Path of the Structural Properties of Multiple Populations in Globular Clusters}",
      journal = {\apjl},
         year = 2019,
        month = oct,
       volume = {884},
       number = {1},
          eid = {L24},
        pages = {L24},
          doi = {10.3847/2041-8213/ab45f7},
archivePrefix = {arXiv},
       eprint = {1910.00613},
 primaryClass = {astro-ph.SR},
       adsurl = {https://ui.adsabs.harvard.edu/abs/2019ApJ...884L..24D}
}

@ARTICLE{Alessandrini2016,
       author = {{Alessandrini}, Emiliano and {Lanzoni}, Barbara and {Ferraro}, Francesco R. and {Miocchi}, Paolo and {Vesperini}, Enrico},
        title = "{Investigating the Mass Segregation Process in Globular Clusters with Blue Straggler Stars: The Impact of Dark Remnants}",
      journal = {\apj},
         year = 2016,
        month = dec,
       volume = {833},
       number = {2},
          eid = {252},
        pages = {252},
          doi = {10.3847/1538-4357/833/2/252},
archivePrefix = {arXiv},
       eprint = {1610.04562},
 primaryClass = {astro-ph.GA},
       adsurl = {https://ui.adsabs.harvard.edu/abs/2016ApJ...833..252A}
}

@ARTICLE{Kuepper2011,
       author = {{K{\"u}pper}, Andreas H.~W. and {Maschberger}, Thomas and {Kroupa}, Pavel and {Baumgardt}, Holger},
        title = "{Mass segregation and fractal substructure in young massive clusters - I. The McLuster code and method calibration}",
      journal = {\mnras},
         year = 2011,
        month = nov,
       volume = {417},
       number = {3},
        pages = {2300-2317},
          doi = {10.1111/j.1365-2966.2011.19412.x},
archivePrefix = {arXiv},
       eprint = {1107.2395},
 primaryClass = {astro-ph.GA},
       adsurl = {https://ui.adsabs.harvard.edu/abs/2011MNRAS.417.2300K}
}

@ARTICLE{mcluster2022,
       author = {{Leveque}, A. and {Giersz}, M. and {Banerjee}, S. and {Vesperini}, E. and {Hong}, J. and {Portegies Zwart}, S.},
        title = "{A Monte Carlo study of early gas expulsion and evolution of star clusters: new simulations with the MOCCA code in the AMUSE framework}",
      journal = {\mnras},
         year = 2022,
        month = aug,
       volume = {514},
       number = {4},
        pages = {5739-5750},
          doi = {10.1093/mnras/stac1690},
archivePrefix = {arXiv},
       eprint = {2206.03404},
 primaryClass = {astro-ph.GA},
       adsurl = {https://ui.adsabs.harvard.edu/abs/2022MNRAS.514.5739L}
}

@ARTICLE{HSB2016,
       author = {{Huang}, Si-Yi and {Spurzem}, Rainer and {Berczik}, Peter},
        title = "{Performance analysis of parallel gravitational N-body codes on large GPU clusters}",
      journal = {Research in Astronomy and Astrophysics},
         year = 2016,
        month = jan,
       volume = {16},
       number = {1},
          eid = {11},
        pages = {11},
          doi = {10.1088/1674-4527/16/1/011},
archivePrefix = {arXiv},
       eprint = {1508.02510},
 primaryClass = {astro-ph.IM},
       adsurl = {https://ui.adsabs.harvard.edu/abs/2016RAA....16...11H}
}

@ARTICLE{Bastian2018,
       author = {{Bastian}, Nate and {Lardo}, Carmela},
        title = "{Multiple Stellar Populations in Globular Clusters}",
      journal = {\araa},
         year = 2018,
        month = sep,
       volume = {56},
        pages = {83-136},
          doi = {10.1146/annurev-astro-081817-051839},
archivePrefix = {arXiv},
       eprint = {1712.01286},
 primaryClass = {astro-ph.SR},
       adsurl = {https://ui.adsabs.harvard.edu/abs/2018ARA&A..56...83B}
}

@ARTICLE{Gratton2019,
       author = {{Gratton}, Raffaele and {Bragaglia}, Angela and {Carretta}, Eugenio and {D'Orazi}, Valentina and {Lucatello}, Sara and {Sollima}, Antonio},
        title = "{What is a globular cluster? An observational perspective}",
      journal = {\aapr},
         year = 2019,
        month = nov,
       volume = {27},
       number = {1},
          eid = {8},
        pages = {8},
          doi = {10.1007/s00159-019-0119-3},
archivePrefix = {arXiv},
       eprint = {1911.02835},
 primaryClass = {astro-ph.SR},
       adsurl = {https://ui.adsabs.harvard.edu/abs/2019A&ARv..27....8G}
}

@ARTICLE{Milone2022,
       author = {{Milone}, Antonino P. and {Marino}, Anna F.},
        title = "{Multiple Populations in Star Clusters}",
      journal = {Universe},
         year = 2022,
        month = jun,
       volume = {8},
       number = {7},
          eid = {359},
        pages = {359},
          doi = {10.3390/universe8070359},
archivePrefix = {arXiv},
       eprint = {2206.10564},
 primaryClass = {astro-ph.GA},
       adsurl = {https://ui.adsabs.harvard.edu/abs/2022Univ....8..359M}
}

@ARTICLE{Leitingeretal2024,
       author = {{Leitinger}, E.~I. and {Baumgardt}, H. and {Cabrera-Ziri}, I. and {Hilker}, M. and {Carbajo-Hijarrubia}, J. and {Gieles}, M. and {Husser}, T.~O. and {Kamann}, S.},
        title = "{The kinematics of 30 Milky Way globular clusters and the multiple stellar populations within}",
      journal = {\aap},
         year = 2025,
        month = feb,
       volume = {694},
          eid = {A184},
        pages = {A184},
          doi = {10.1051/0004-6361/202452477},
archivePrefix = {arXiv},
       eprint = {2410.02855},
 primaryClass = {astro-ph.GA},
       adsurl = {https://ui.adsabs.harvard.edu/abs/2025A&A...694A.184L}
}

@ARTICLE{Leitingeretal2023,
       author = {{Leitinger}, E. and {Baumgardt}, H. and {Cabrera-Ziri}, I. and {Hilker}, M. and {Pancino}, E.},
        title = "{A wide-field view on multiple stellar populations in 28 Milky Way globular clusters}",
      journal = {\mnras},
         year = 2023,
        month = mar,
       volume = {520},
       number = {1},
        pages = {1456-1480},
          doi = {10.1093/mnras/stad093},
archivePrefix = {arXiv},
       eprint = {2301.04166},
 primaryClass = {astro-ph.GA},
       adsurl = {https://ui.adsabs.harvard.edu/abs/2023MNRAS.520.1456L}
}

@ARTICLE{Gierszetal2025,
       author = {{Giersz}, M. and {Askar}, A. and {Hypki}, A. and {Hong}, J. and {Wiktorowicz}, G. and {Hellstr{\"o}m}, L.},
        title = "{Multiple stellar populations in MOCCA globular cluster models: Transient spatial overconcentration of pristine red giant stars driven by strong dynamical encounters}",
      journal = {\aap},
         year = 2025,
        month = jun,
       volume = {698},
          eid = {L11},
        pages = {L11},
          doi = {10.1051/0004-6361/202554233},
archivePrefix = {arXiv},
       eprint = {2502.17162},
 primaryClass = {astro-ph.GA},
       adsurl = {https://ui.adsabs.harvard.edu/abs/2025A&A...698L..11G}
}

@ARTICLE{Hypki2013,
       author = {{Hypki}, Arkadiusz and {Giersz}, Mirek},
        title = "{MOCCA code for star cluster simulations - I. Blue stragglers, first results}",
      journal = {\mnras},
         year = "2013",
        month = "Feb",
       volume = {429},
       number = {2},
        pages = {1221-1243},
          doi = {10.1093/mnras/sts415},
archivePrefix = {arXiv},
       eprint = {1207.6700},
 primaryClass = {astro-ph.GA},
       adsurl = {https://ui.adsabs.harvard.edu/abs/2013MNRAS.429.1221H}
}

@ARTICLE{Gierszetal2013,
       author = {{Giersz}, Mirek and {Heggie}, Douglas C. and {Hurley}, Jarrod R. and
         {Hypki}, Arkadiusz},
        title = "{MOCCA code for star cluster simulations - II. Comparison with N-body simulations}",
      journal = {\mnras},
         year = "2013",
        month = "May",
       volume = {431},
       number = {3},
        pages = {2184-2199},
          doi = {10.1093/mnras/stt307},
archivePrefix = {arXiv},
       eprint = {1112.6246},
 primaryClass = {astro-ph.GA},
       adsurl = {https://ui.adsabs.harvard.edu/abs/2013MNRAS.431.2184G}
}

@ARTICLE{Giersz1998,
       author = {{Giersz}, Mirek},
        title = "{Monte Carlo simulations of star clusters - I. First Results}",
      journal = {\mnras},
         year = 1998,
        month = aug,
       volume = {298},
       number = {4},
        pages = {1239-1248},
          doi = {10.1046/j.1365-8711.1998.01734.x},
archivePrefix = {arXiv},
       eprint = {astro-ph/9804127},
 primaryClass = {astro-ph},
       adsurl = {https://ui.adsabs.harvard.edu/abs/1998MNRAS.298.1239G}
}

@ARTICLE{Hypki2022,
       author = {{Hypki}, Arkadiusz and {Giersz}, Mirek and {Hong}, Jongsuk and {Leveque}, Agostino and {Askar}, Abbas and {Belloni}, Diogo and {Otulakowska-Hypka}, Magdalena},
        title = "{MOCCA: dynamics and evolution of single and binary stars of multiple stellar populations in tidally filling and underfilling globular star clusters}",
      journal = {\mnras},
         year = 2022,
        month = dec,
       volume = {517},
       number = {4},
        pages = {4768-4787},
          doi = {10.1093/mnras/stac2815},
archivePrefix = {arXiv},
       eprint = {2205.05397},
 primaryClass = {astro-ph.GA},
       adsurl = {https://ui.adsabs.harvard.edu/abs/2022MNRAS.517.4768H}
}

@ARTICLE{Hypki2025,
       author = {{Hypki}, A. and {Vesperini}, E. and {Giersz}, M. and {Hong}, J. and {Askar}, A. and {Otulakowska-Hypka}, M. and {Hellstrom}, L. and {Wiktorowicz}, G.},
        title = "{MOCCA: Global properties of tidally filling and underfilling globular star clusters with multiple stellar populations}",
      journal = {\aap},
         year = 2025,
        month = jan,
       volume = {693},
          eid = {A41},
        pages = {A41},
          doi = {10.1051/0004-6361/202348653},
archivePrefix = {arXiv},
       eprint = {2406.08059},
 primaryClass = {astro-ph.GA},
       adsurl = {https://ui.adsabs.harvard.edu/abs/2025A&A...693A..41H}
}

@ARTICLE{Henon1971,
       author = {{H{\'e}non}, M.~H.},
        title = "{The Monte Carlo Method (Papers appear in the Proceedings of IAU Colloquium No. 10 Gravitational N-Body Problem (ed. by Myron Lecar), R. Reidel Publ. Co. , Dordrecht-Holland.)}",
      journal = {\apss},
         year = 1971,
        month = nov,
       volume = {14},
       number = {1},
        pages = {151-167},
          doi = {10.1007/BF00649201},
       adsurl = {https://ui.adsabs.harvard.edu/abs/1971Ap&SS..14..151H}
}

@ARTICLE{Stodolkiewicz1982,
       author = {{Stod{\'o}{\l}kiewicz}, J.~S.},
        title = "{Dynamical evolution of globular clusters. I}",
      journal = {\actaa},
         year = 1982,
        month = jan,
       volume = {32},
       number = {1-2},
        pages = {63-91},
       adsurl = {https://ui.adsabs.harvard.edu/abs/1982AcA....32...63S}
}

@ARTICLE{Fregeauetal2004,
       author = {{Fregeau}, J.~M. and {Cheung}, P. and {Portegies Zwart}, S.~F. and {Rasio}, F.~A.},
        title = "{Stellar collisions during binary-binary and binary-single star interactions}",
      journal = {\mnras},
         year = 2004,
        month = jul,
       volume = {352},
       number = {1},
        pages = {1-19},
          doi = {10.1111/j.1365-2966.2004.07914.x},
archivePrefix = {arXiv},
       eprint = {astro-ph/0401004},
 primaryClass = {astro-ph},
       adsurl = {https://ui.adsabs.harvard.edu/abs/2004MNRAS.352....1F}
}

@ARTICLE{Fregeau2007,
       author = {{Fregeau}, John M. and {Rasio}, Frederic A.},
        title = "{Monte Carlo Simulations of Globular Cluster Evolution. IV. Direct Integration of Strong Interactions}",
      journal = {\apj},
         year = 2007,
        month = apr,
       volume = {658},
       number = {2},
        pages = {1047-1061},
          doi = {10.1086/511809},
archivePrefix = {arXiv},
       eprint = {astro-ph/0608261},
 primaryClass = {astro-ph},
       adsurl = {https://ui.adsabs.harvard.edu/abs/2007ApJ...658.1047F}
}

@ARTICLE{Kamlahetal2022,
       author = {{Kamlah}, A.~W.~H. and {Leveque}, A. and {Spurzem}, R. and {Arca Sedda}, M. and {Askar}, A. and {Banerjee}, S. and {Berczik}, P. and {Giersz}, M. and {Hurley}, J. and {Belloni}, D. and {K{\"u}hmichel}, L. and {Wang}, L.},
        title = "{Preparing the next gravitational million-body simulations: evolution of single and binary stars in NBODY6++GPU, MOCCA, and MCLUSTER}",
      journal = {\mnras},
         year = 2022,
        month = apr,
       volume = {511},
       number = {3},
        pages = {4060-4089},
          doi = {10.1093/mnras/stab3748},
archivePrefix = {arXiv},
       eprint = {2105.08067},
 primaryClass = {astro-ph.GA},
       adsurl = {https://ui.adsabs.harvard.edu/abs/2022MNRAS.511.4060K}
}

@ARTICLE{Gierszetal2025b,
       author = {{Giersz}, M. and {Askar}, A. and {Hypki}, A. and {Hong}, J. and {Wiktorowicz}, G. and {Hellstr{\"o}m}, L.},
        title = "{MOCCA: Effects of pristine gas accretion and cluster migration on globular cluster evolution, global parameters, and multiple stellar populations}",
      journal = {\aap},
         year = 2025,
        month = jul,
       volume = {699},
          eid = {A76},
        pages = {A76},
          doi = {10.1051/0004-6361/202452945},
archivePrefix = {arXiv},
       eprint = {2411.06421},
 primaryClass = {astro-ph.GA},
       adsurl = {https://ui.adsabs.harvard.edu/abs/2025A&A...699A..76G}
}

@ARTICLE{Giersz2008,
       author = {{Giersz}, Mirek and {Heggie}, Douglas C. and {Hurley}, Jarrod R.},
        title = "{Monte Carlo simulations of star clusters - IV. Calibration of the Monte Carlo code and comparison with observations for the open cluster M67}",
      journal = {\mnras},
         year = 2008,
        month = jul,
       volume = {388},
       number = {1},
        pages = {429-443},
          doi = {10.1111/j.1365-2966.2008.13407.x},
archivePrefix = {arXiv},
       eprint = {0801.3968},
 primaryClass = {astro-ph},
       adsurl = {https://ui.adsabs.harvard.edu/abs/2008MNRAS.388..429G}
}

@ARTICLE{Wang2016,
       author = {{Wang}, Long and {Spurzem}, Rainer and {Aarseth}, Sverre and {Giersz}, Mirek and {Askar}, Abbas and {Berczik}, Peter and {Naab}, Thorsten and {Schadow}, Riko and {Kouwenhoven}, M.~B.~N.},
        title = "{The DRAGON simulations: globular cluster evolution with a million stars}",
      journal = {\mnras},
         year = 2016,
        month = may,
       volume = {458},
       number = {2},
        pages = {1450-1465},
          doi = {10.1093/mnras/stw274},
archivePrefix = {arXiv},
       eprint = {1602.00759},
 primaryClass = {astro-ph.SR},
       adsurl = {https://ui.adsabs.harvard.edu/abs/2016MNRAS.458.1450W}
}

@ARTICLE{Madrid2017,
       author = {{Madrid}, Juan P. and {Leigh}, Nathan W.~C. and {Hurley}, Jarrod R. and {Giersz}, Mirek},
        title = "{Mass evaporation rate of globular clusters in a strong tidal field}",
      journal = {\mnras},
         year = 2017,
        month = sep,
       volume = {470},
       number = {2},
        pages = {1729-1737},
          doi = {10.1093/mnras/stx1350},
archivePrefix = {arXiv},
       eprint = {1706.06635},
 primaryClass = {astro-ph.GA},
       adsurl = {https://ui.adsabs.harvard.edu/abs/2017MNRAS.470.1729M}
}

@ARTICLE{Geller2019,
       author = {{Geller}, Aaron M. and {Leigh}, Nathan W.~C. and {Giersz}, Mirek and {Kremer}, Kyle and {Rasio}, Frederic A.},
        title = "{In Search of the Thermal Eccentricity Distribution}",
      journal = {\apj},
         year = 2019,
        month = feb,
       volume = {872},
       number = {2},
          eid = {165},
        pages = {165},
          doi = {10.3847/1538-4357/ab0214},
archivePrefix = {arXiv},
       eprint = {1902.00019},
 primaryClass = {astro-ph.SR},
       adsurl = {https://ui.adsabs.harvard.edu/abs/2019ApJ...872..165G}
}

@ARTICLE{Rizzuto2021,
       author = {{Rizzuto}, Francesco Paolo and {Naab}, Thorsten and {Spurzem}, Rainer and {Giersz}, Mirek and {Ostriker}, J.~P. and {Stone}, N.~C. and {Wang}, Long and {Berczik}, Peter and {Rampp}, M.},
        title = "{Intermediate mass black hole formation in compact young massive star clusters}",
      journal = {\mnras},
         year = 2021,
        month = mar,
       volume = {501},
       number = {4},
        pages = {5257-5273},
          doi = {10.1093/mnras/staa3634},
archivePrefix = {arXiv},
       eprint = {2008.09571},
 primaryClass = {astro-ph.GA},
       adsurl = {https://ui.adsabs.harvard.edu/abs/2021MNRAS.501.5257R}
}

@ARTICLE{Vergara2025,
       author = {{Vergara}, Marcelo C. and {Askar}, Abbas and {Kamlah}, Albrecht W.~H. and {Spurzem}, Rainer and {Flammini Dotti}, Francesco and {Schleicher}, Dominik R.~G. and {Arca Sedda}, Manuel and {Hypki}, Arkadiusz and {Giersz}, Mirek and {Hurley}, Jarrod and {Berczik}, Peter and {Escala}, Andres and {Hoyer}, Nils and {Neumayer}, Nadine and {Pang}, Xiaoying and {Tanikawa}, Ataru and {Cen}, Renyue and {Naab}, Thorsten},
        title = "{Rapid formation of a very massive star (>50000 M$_{{\ensuremath{\odot}}}$), and subsequently, of an IMBH, from runaway collisions: Direct N-body and Monte Carlo simulations of dense star clusters}",
      journal = {\aap},
         year = 2025,
        month = dec,
       volume = {704},
          eid = {A321},
        pages = {A321},
          doi = {10.1051/0004-6361/202555307},
archivePrefix = {arXiv},
       eprint = {2505.07491},
 primaryClass = {astro-ph.GA},
       adsurl = {https://ui.adsabs.harvard.edu/abs/2025A&A...704A.321V}
}

@ARTICLE{Kroupa2001,
       author = {{Kroupa}, Pavel},
        title = "{On the variation of the initial mass function}",
      journal = {\mnras},
         year = 2001,
        month = apr,
       volume = {322},
       number = {2},
        pages = {231-246},
          doi = {10.1046/j.1365-8711.2001.04022.x},
archivePrefix = {arXiv},
       eprint = {astro-ph/0009005},
 primaryClass = {astro-ph},
       adsurl = {https://ui.adsabs.harvard.edu/abs/2001MNRAS.322..231K}
}

@ARTICLE{King1966,
       author = {{King}, Ivan R.},
        title = "{The structure of star clusters. III. Some simple dynamical models}",
      journal = {\aj},
         year = 1966,
        month = feb,
       volume = {71},
        pages = {64},
          doi = {10.1086/109857},
       adsurl = {https://ui.adsabs.harvard.edu/abs/1966AJ.....71...64K}
}

@ARTICLE{Baumgardt2003,
       author = {{Baumgardt}, Holger and {Makino}, Junichiro},
        title = "{Dynamical evolution of star clusters in tidal fields}",
      journal = {\mnras},
         year = 2003,
        month = mar,
       volume = {340},
       number = {1},
        pages = {227-246},
          doi = {10.1046/j.1365-8711.2003.06286.x},
archivePrefix = {arXiv},
       eprint = {astro-ph/0211471},
 primaryClass = {astro-ph},
       adsurl = {https://ui.adsabs.harvard.edu/abs/2003MNRAS.340..227B}
}

@ARTICLE{Spurzem2023,
       author = {{Spurzem}, Rainer and {Kamlah}, Albrecht},
        title = "{Computational methods for collisional stellar systems}",
      journal = {Living Reviews in Computational Astrophysics},
         year = 2023,
        month = dec,
       volume = {9},
       number = {1},
          eid = {3},
        pages = {3},
          doi = {10.1007/s41115-023-00018-w},
archivePrefix = {arXiv},
       eprint = {2305.11606},
 primaryClass = {astro-ph.IM},
       adsurl = {https://ui.adsabs.harvard.edu/abs/2023LRCA....9....3S}
}

@ARTICLE{Wangetal2015,
       author = {{Wang}, Long and {Spurzem}, Rainer and {Aarseth}, Sverre and {Nitadori}, Keigo and {Berczik}, Peter and et al.},
        title = "{NBODY6++GPU: ready for the gravitational million-body problem}",
      journal = {MNRAS},
         year = 2015,
        month = jul,
       volume = {450},
       number = {4},
        pages = {4070-4080},
          doi = {10.1093/mnras/stv817},
archivePrefix = {arXiv},
       eprint = {1504.03687},
 primaryClass = {astro-ph.IM},
       adsurl = {https://ui.adsabs.harvard.edu/abs/2015MNRAS.450.4070W}
}

@Article{nitadori2012,
  author        = {{Nitadori}, Keigo and {Aarseth}, Sverre J.},
  journal       = {\mnras},
  title         = {{Accelerating NBODY6 with graphics processing units}},
  year          = {2012},
  month         = {Jul},
  number        = {1},
  pages         = {545-552},
  volume        = {424},
  adsurl        = {https://ui.adsabs.harvard.edu/abs/2012MNRAS.424..545N},
  archiveprefix = {arXiv},
  doi           = {10.1111/j.1365-2966.2012.21227.x},
  eprint        = {1205.1222},
  primaryclass  = {astro-ph.IM},
}

@Article{kustaanheimo1965,
  author        = {Kustaanheimo, P. and Stiefel, E.},
  journal       = {Journal fur die reine und angewandte Mathematik},
  title         = {Perturbation theory of Kepler motion based on spinor regularization.},
  year          = {1965},
  number        = {218},
}

@ARTICLE{Cadelano2024,
       author = {{Cadelano}, Mario and {Dalessandro}, Emanuele and {Vesperini}, Enrico},
        title = "{The structural properties of multiple populations in globular clusters: The instructive case of NGC 3201}",
      journal = {\aap},
         year = 2024,
        month = may,
       volume = {685},
          eid = {A158},
        pages = {A158},
          doi = {10.1051/0004-6361/202349021},
archivePrefix = {arXiv},
       eprint = {2402.09514},
 primaryClass = {astro-ph.GA},
       adsurl = {https://ui.adsabs.harvard.edu/abs/2024A&A...685A.158C}
}

@ARTICLE{Mehta2025,
       author = {{Mehta}, V.~J. and {Milone}, A.~P. and {Casagrande}, L. and {Marino}, A.~F. and {Legnardi}, M.~V. and {Cordoni}, G. and {Dondoglio}, E. and {Jang}, S. and {Lionetto}, S. and {Ziliotto}, T. and {Barbieri}, M. and {Bernizzoni}, M. and {Bortolan}, E. and {Bouras Moreno Sanchez}, A. and {Lagioia}, E.~P. and {Mohandasan}, A. and {Muratore}, F.},
        title = "{Spectrophotometry and radial distribution of multiple stellar populations in globular clusters from Gaia XP spectra}",
      journal = {\mnras},
         year = 2025,
        month = jan,
       volume = {536},
       number = {2},
        pages = {1077-1088},
          doi = {10.1093/mnras/stae2542},
archivePrefix = {arXiv},
       eprint = {2406.02755},
 primaryClass = {astro-ph.SR},
       adsurl = {https://ui.adsabs.harvard.edu/abs/2025MNRAS.536.1077M}
}

@ARTICLE{Dalessandro2024,
       author = {{Dalessandro}, E. and {Cadelano}, M. and {Della Croce}, A. and {Aros}, F.~I. and {White}, E.~B. and {Vesperini}, E. and {Fanelli}, C. and {Ferraro}, F.~R. and {Lanzoni}, B. and {Leanza}, S. and {Origlia}, L.},
        title = "{A 3D view of multiple populations' kinematics in Galactic globular clusters}",
      journal = {\aap},
         year = 2024,
        month = nov,
       volume = {691},
          eid = {A94},
        pages = {A94},
          doi = {10.1051/0004-6361/202451054},
archivePrefix = {arXiv},
       eprint = {2409.03827},
 primaryClass = {astro-ph.GA},
       adsurl = {https://ui.adsabs.harvard.edu/abs/2024A&A...691A..94D}
}

@ARTICLE{Libralato2023,
       author = {{Libralato}, Mattia and {Vesperini}, Enrico and {Bellini}, Andrea and {Milone}, Antonino P. and {van der Marel}, Roeland P. and {Piotto}, Giampaolo and {Anderson}, Jay and {Aparicio}, Antonio and {Barbuy}, Beatriz and {Bedin}, Luigi R. and {Brown}, Thomas M. and {Cassisi}, Santi and {Nardiello}, Domenico and {Sarajedini}, Ata and {Scalco}, Michele},
        title = "{The Hubble Space Telescope UV Legacy Survey of Galactic Globular Clusters. XXIV. Differences in Internal Kinematics of Multiple Stellar Populations}",
      journal = {\apj},
         year = 2023,
        month = feb,
       volume = {944},
       number = {1},
          eid = {58},
        pages = {58},
          doi = {10.3847/1538-4357/acaec6},
archivePrefix = {arXiv},
       eprint = {2301.04148},
 primaryClass = {astro-ph.GA},
       adsurl = {https://ui.adsabs.harvard.edu/abs/2023ApJ...944...58L}
}

@ARTICLE{Cordoni2025,
       author = {{Cordoni}, G. and {Casagrande}, L. and {Milone}, A.~P. and {Dondoglio}, E. and {Mastrobuono-Battisti}, A. and {Jang}, S. and {Marino}, A.~F. and {Lagioia}, E.~P. and {Legnardi}, M.~V. and {Ziliotto}, T. and {Muratore}, F. and {Mehta}, V. and {Lacchin}, E. and {Tailo}, M.},
        title = "{Internal dynamics of multiple populations in 28 Galactic globular clusters: a wide-field study with Gaia and the Hubble Space Telescope}",
      journal = {\mnras},
         year = 2025,
        month = mar,
       volume = {537},
       number = {3},
        pages = {2342-2361},
          doi = {10.1093/mnras/staf102},
archivePrefix = {arXiv},
       eprint = {2409.02330},
 primaryClass = {astro-ph.GA},
       adsurl = {https://ui.adsabs.harvard.edu/abs/2025MNRAS.537.2342C}
}

@ARTICLE{Kamann2020,
       author = {{Kamann}, S. and {Dalessandro}, E. and {Bastian}, N. and {Brinchmann}, J. and {den Brok}, M. and {Dreizler}, S. and {Giesers}, B. and {G{\"o}ttgens}, F. and {Husser}, T.-O. and {Krajnovi{\'c}}, D. and {van de Ven}, G. and {Watkins}, L.~L. and {Wisotzki}, L.},
        title = "{The peculiar kinematics of the multiple populations in the globular cluster Messier 80 (NGC 6093)}",
      journal = {\mnras},
         year = 2020,
        month = feb,
       volume = {492},
       number = {1},
        pages = {966-977},
          doi = {10.1093/mnras/stz3506},
archivePrefix = {arXiv},
       eprint = {1912.06158},
 primaryClass = {astro-ph.SR},
       adsurl = {https://ui.adsabs.harvard.edu/abs/2020MNRAS.492..966K}
}

@ARTICLE{Kamann2020b,
       author = {{Kamann}, S. and {Giesers}, B. and {Bastian}, N. and {Brinchmann}, J. and {Dreizler}, S. and {G{\"o}ttgens}, F. and {Husser}, T.-O. and {Latour}, M. and {Weilbacher}, P.~M. and {Wisotzki}, L.},
        title = "{The binary content of multiple populations in NGC 3201}",
      journal = {\aap},
         year = 2020,
        month = mar,
       volume = {635},
          eid = {A65},
        pages = {A65},
          doi = {10.1051/0004-6361/201936843},
archivePrefix = {arXiv},
       eprint = {1912.01627},
 primaryClass = {astro-ph.SR},
       adsurl = {https://ui.adsabs.harvard.edu/abs/2020A&A...635A..65K}
}

@ARTICLE{pavlik2025,
       author = {{Pavl{\'\i}k}, V{\'a}clav and {Davies}, Melvyn B. and {Leitinger}, Ellen I. and {Baumgardt}, Holger and {Bobrick}, Alexey and {Cabrera-Ziri}, Ivan and {Hilker}, Michael and {Winter}, Andrew J.},
        title = "{Spatial mixing of stellar populations in globular clusters via binary─single star scattering}",
      journal = {\aap},
         year = 2025,
        month = nov,
       volume = {703},
          eid = {A157},
        pages = {A157},
          doi = {10.1051/0004-6361/202556753},
archivePrefix = {arXiv},
       eprint = {2508.03322},
 primaryClass = {astro-ph.GA},
       adsurl = {https://ui.adsabs.harvard.edu/abs/2025A&A...703A.157P}
}

@ARTICLE{LiSpurzem2026,
       author = {{Li}, Zhong-Mu and {Spurzem}, Rainer},
        title = "{NbodyCP: A Direct N-body Simulation Code for Composite Stellar Populations of Single and Binary Star Clusters}",
      journal = {Research in Astronomy and Astrophysics},
         year = 2026,
        month = jan,
       volume = {26},
       number = {1},
          eid = {015009},
        pages = {015009},
          doi = {10.1088/1674-4527/ae1ae8},
archivePrefix = {arXiv},
       eprint = {2511.06181},
 primaryClass = {astro-ph.GA},
       adsurl = {https://ui.adsabs.harvard.edu/abs/2026RAA....26a5009L}
}



\begin{appendix}

\onecolumn 


\section{The time evolution of the A$^+$ parameter for the evolved 
luminous stars (EVL) and main sequence (MS) stars}
\label{sec:app1}

\begin{figure*}[hpbt!]
\centering
\includegraphics[width=0.45\linewidth]{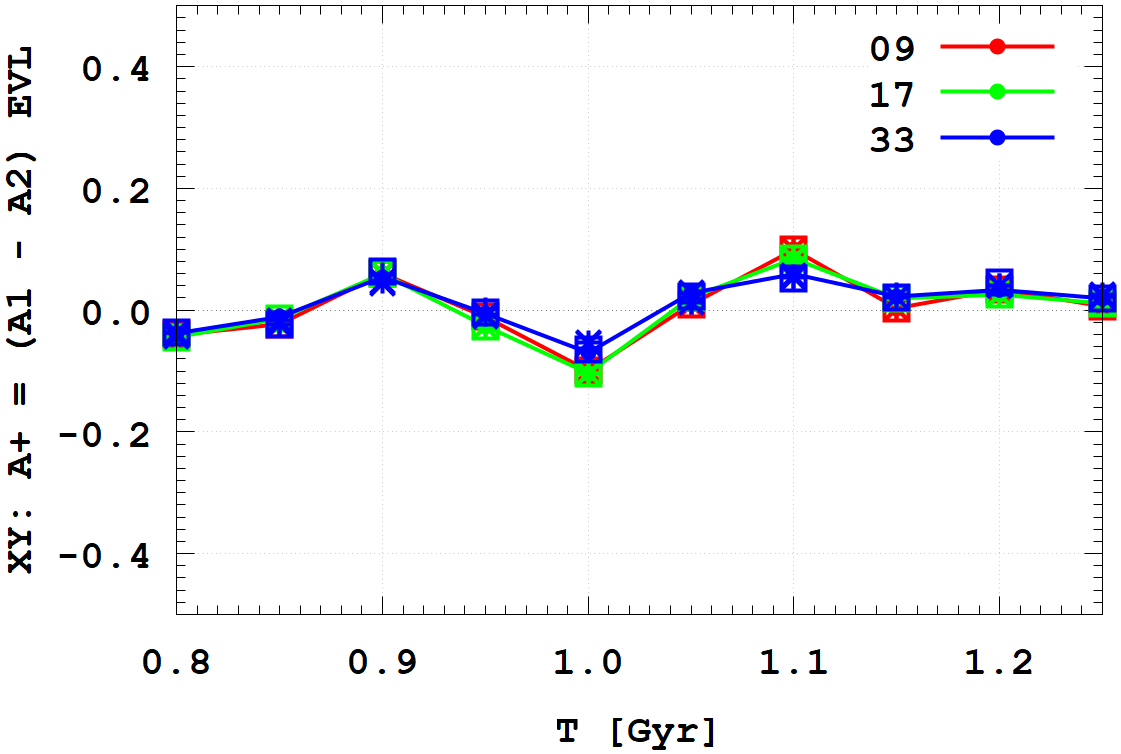}
\includegraphics[width=0.45\linewidth]{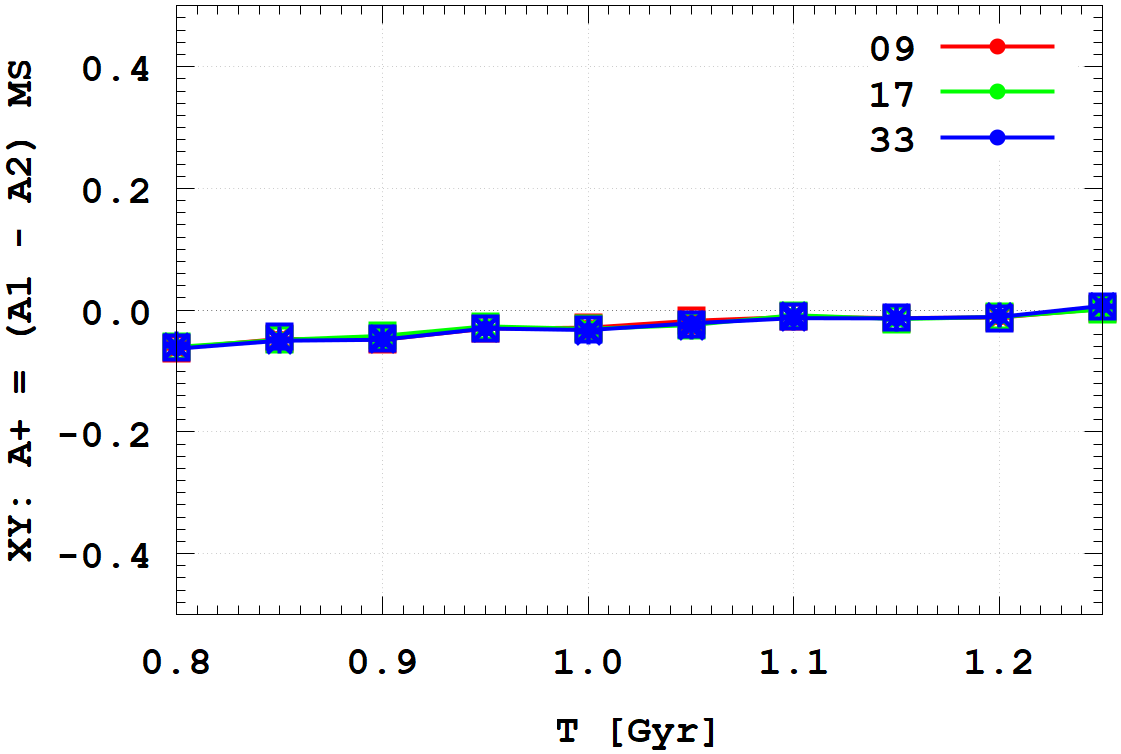}
\caption{The time evolution of the A$^+$ parameter for different type of stars and for $X-Y$ projection of 3D snapshots. The left panel shows the evolved luminous stars, and the right panel -- MS stars for 5 different random selections, which also reduce the star number to the level of RGB stars at each snapshot. The red, green and blue lines are averaged over 9, 17 and 33 snapshots, respectively.}
\label{fig:def-Aplus-01-sel}
\end{figure*}

In Fig.~\ref{fig:def-Aplus-01-sel} we specially check the time evolution of the A$^+$ parameter 
for the evolved luminous stars (EVL) and main sequence (MS) stars but in 5 different randomly 
selected samples in such a way that the number of sampled stars almost equal to the much lower 
number of corresponding in time RGB stars in both populations. As we can see, independent of 
the randomisation and besides a much lower number of stars compared to the data on 
Fig.~\ref{fig:def-Aplus-01}, we see very similar A$^+$ parameter values. Also, we see that 
these values for the EVL and MS stars are much smaller compared to the original RGB stars data.

\end{appendix}

\end{document}